\documentclass[12pt]{article}
\usepackage{epsfig}
\usepackage{slashed}
\usepackage{listings}
\usepackage{booktabs,multirow,tabularx}
\usepackage{amsmath}
\usepackage{float} 
\usepackage{subfig} 

\newcommand{\ben}{\begin{enumerate}}
\newcommand{\een}{\end{enumerate}}
\newcommand{\bit}{\begin{itemize}}
\newcommand{\eit}{\end{itemize}}
\newcommand{\bc}{\begin{center}}
\newcommand{\ec}{\end{center}}
\newcommand{\bq}{\begin{equation}}
\newcommand{\eq}{\end{equation}}
\newcommand{\bqa}{\begin{eqnarray}}
\newcommand{\eqa}{\end{eqnarray}}

\def\jpsi{{J/\psi}}

\def\ss{{\bigl.^3\hspace{-1mm}S^{[1]}_1}}

\def\p0{{\bigl.^3\hspace{-1mm}P^{[8]}_0}}

\def\tsos{{\bigl.^3\hspace{-1mm}S^{[1]}_1}}
\def\oszo{{\bigl.^1\hspace{-1mm}S^{[8]}_0}}
\def\oszs{{\bigl.^1\hspace{-1mm}S^{[1]}_0}}
\def\tsoo{{\bigl.^3\hspace{-1mm}S^{[8]}_1}}

\def\tpjst{{\bigl.^3\hspace{-1mm}P^{[1]}_{J=0,1,2}}}
\def\tpjot{{\bigl.^3\hspace{-1mm}P^{[8]}_{J=0,1,2}}}
\def\opos{{\bigl.^1\hspace{-1mm}P^{[1]}_1}}
\def\opoo{{\bigl.^1\hspace{-1mm}P^{[8]}_1}}

\def\oszo{{\bigl.^1\hspace{-1mm}S^{[8]}_0}}
\def\tsoo{{\bigl.^3\hspace{-1mm}S^{[8]}_1}}

\def\to{\rightarrow}

\def\HELACOnia{ {\sc \small HELAC-Onia}}
\def\HELAC{ {\sc \small HELAC}}
\def\PHEGAS{{\sc\small PHEGAS}}
\def\Pythia{{\sc\small Pythia}}
\def\Herwig{{\sc\small Herwig}}
\def\QEDPS{{\sc\small QEDPS}}
\def\Python{{\sc \small Python}}
\def\Fortran{{\sc \small Fortran}}
\def\Cpp{{\sc \small C++}}
\def\Unix{{\sc \small Unix}}
\def\Minuit{{\sc\small Minuit}}
\def\MG5aMC{{\sc \small MadGraph5\_aMC@NLO}}
\def\MadGraph{{\sc \small MadGraph}}
\def\FastJet{{\sc \small FastJet}}
\def\FJCore{{\sc \small FJCore}}
\def\HEPTopTagger{{\sc \small HEPTopTagger}}
\def\HepMC{{\sc \small HepMC}}
\def\TopDrawer{{\sc \small TopDrawer}}
\def\Hbook{{\sc \small Hbook}}
\def\Root{{\sc \small Root}}
\def\Mint{{\sc \small Mint}}
\def\Rambo{{\sc \small Rambo}}
\def\Gnuplot{{\sc \small Gnuplot}}
\def\Ranmar{{\sc \small Ranmar}}
\def\MadOnia{{\sc \small MadOnia}}
\def\EvtGen{{\sc \small EvtGen}}
\def\cCode#1{\begin{lstlisting}[mathescape,basicstyle=\small
\ttfamily,frame=leftline,aboveskip=4mm,belowskip=4mm,xleftmargin=20pt,framexleftmargin=10pt,
numbers=none,framerule=2pt,abovecaptionskip=0.0mm,belowcaptionskip=3.5mm #1]}
\def\chCode#1{\begin{lstlisting}[mathescape=true,basicstyle=\small
\ttfamily]}

\begin{document}

\pagestyle{empty}

\begin{flushright}
CERN-PH-TH-2015-155 \\
\end{flushright}

\vspace*{2cm}

\bc\begin{LARGE} {\bf HELAC-Onia 2.0: an upgraded matrix-element and event generator for heavy quarkonium physics}\\
\end{LARGE} \vspace*{2cm}

{\large {\bf Hua-Sheng Shao} 
} \\[12pt]
 {
 ~PH Department, TH Unit, CERN, CH-1211, Geneva 23, Switzerland}
\\{E-mail:huasheng.shao@cern.ch}

\vspace*{1cm}

{\bf ABSTRACT}\\[12pt]  \ec

\begin{quote}

We present an upgraded version (denoted as version 2.0) of the program \HELACOnia\ for the automated computation of heavy-quarkonium helicity amplitudes within non-relativistic QCD framework. The new code has been designed to include many new and useful features for practical phenomenological simulations. It is designed for job submissions under cluster enviroment for parallel computations via \Python\ scripts. We have interfaced \HELACOnia\ to the parton shower Monte Carlo programs \Pythia\ 8 and \QEDPS\ to take into account the parton-shower effects. Moreover, the decay module guarantees that the program can perform the spin-entangled (cascade-)decay of heavy quarkonium after its generation. We have also implemented a reweighting method to automatically estimate the uncertainties from renormalization and/or factorization scales as well as parton-distribution functions to weighted or unweighted events. A futher update is the possiblity to generate one-dimensional or two-dimensional plots encoded in the analysis files on the fly. Some dedicated examples are given at the end of the writeup.

\end{quote}

%
%





\newpage
\pagestyle{plain}


\bc {\bf PROGRAM SUMMARY}\\[18pt]\ec
{\it Program title:} \\
{\tt \HELACOnia\ 2.0}. \\[8pt]
{\it Program obtainable from:}\\
http://helac-phegas.web.cern.ch/helac-phegas\\
[8pt]
{\it Licensing provisions:} none\\[8pt]
{\it Operating system under which the program has been tested:}\\
\Unix-like platform.\\[8pt]
{\it Programming language:} \\
\Python, \Fortran\ 77, \Fortran\ 90, \Cpp\\[8pt]
{\it Keywords:} \\
heavy quarkonium, NRQCD, Monte Carlo simulation
\\[8pt]
{\it Nature of physical problem:}\\
Heavy quarkonium production processes provide an important way to investigate QCD in its poorly known
non-perturbative regime. Its production mechanism  has been attracted extensive interests from the high-energy physics community in decades. The qualitative and quantitative description of heavy-quarkonium production requires complex perturbative computations for high-multiplicity processes in the framework of the well established non-relativistic effecitive theory, NRQCD, and reliable Monte Carlo simulations to repreduce the  collider enviroment.
\\[8pt]
{\it Method of solution:} \\
Based on a recursion relation, the program is able to calculate the helicity ampltiudes of the high-multiplicity heavy-qurkonium-production processes. Several modules are also designed for dedicated simulations: 1) The code has been interfaced with the parton shower Monte Carlo programs; 2) A decay module to let heavy quarkonia decay with correct spin-correlations has been implemented; 3) The code estimates the theoretical uncertainties and analyzes the generated events on the fly; 4) The code is compilant with multi-threading/multi-core usage or cluster processors.
\\[8pt]
{\it CPC classification code:}\\
4.4 Feynman Diagrams, 11.1 General, High Energy Physics and Computing, 11.2 Phase Space and Event Simulation, 11.5 Quantum Chromodynamics, Lattice Gauge Theory
\\[8pt]
{\it Typical running time:}\\
It depends on the process to be calculated and the required accuracy.
\newpage




\section{Introduction}

Since the breakthrough discovery of the Higgs boson at the LHC, much hope has been put on searching beyond Standard Model (BSM) particles in the next runs of LHC. However, the studies of quantum chromodynamics (QCD) are always playing a crucial role in the LHC objectives mainly because QCD is still a poorly-known theory especially in the non-perturbative region due to its color confinement and because it is crucial to understand QCD background at the hadron colliders. For example, at 14 TeV LHC, each bunch crossing will generate around 50 pile-ups. Such an effect is mainly governed by soft interactions. Hence, all aspects of QCD still deserve to be explored as fully as possible.

For a long time, heavy-quarkonium production and decay at high-energy colliders was thought to provide an ideal opporturnity to study both the perturbative and non-perturbative aspects of QCD. Besides, it also shows a rich physics. Fundamental parameters such as the strong coupling constant $\alpha_s$~\cite{Jamin:1997rt,Brambilla:2001qk}, the heavy-flavor quark mass~\cite{Jamin:1997rt,Brambilla:2001qk,Kharchilava:1999yj}, the elements of the Cabibbo-Kobayashi-Maskawa (CKM) matrix~\cite{CKMWeb}, the Yukawa coupling~\cite{Bodwin:2013gca} can be measured with heavy-quarkonium-production processes. Non-perturbative parton-distribution functions--either collinear~\cite{Martin:1987vw,Glover:1987az} or transverse~\cite{Dunnen:2014eta} in the initial hadron--can be constrained from heavy-quarkonium data. B-meson-decay process $B_d^0\to J/\psi+K_s^0$ provides a golden channel to investigate CP violation. Heavy-quarkonium production is also useful in probing the multiple-parton interactions~\cite{Kom:2011bd,Lansberg:2014swa}. Other applications include quarkonium in quark-gluon plasma~\cite{Mocsy:2013syh}, cold nuclear matter effects on quarkonium production~\cite{Ferreiro:2008wc} and even BSM searching~(see e.g. Ref.~\cite{Zhang:2009pt} and references therein) etc.

Despite of its importance, one has very limited choice of Monte Carlo tools for the simulation of the heavy-quarkonium-production processes on the market. From our point of view, this can be attributed to several longstanding puzzles in understanding its mechanism (see e.g Refs.~\cite{Brambilla:2010cs,Andronic:2015wma}) inspite of the well-established effective theory non-relativistic QCD (NRQCD)~\cite{Bodwin:1994jh}. Both \MadOnia~\cite{Artoisenet:2007qm} and \HELACOnia~\cite{Shao:2012iz} are such tools dedicated to matrix-element calculations and event generation within the NRQCD framework, which aim at providing general and user-friendly public tools for theorists and experimentalists to study the quarkonium physics. Although there are many similarities in both tools, we wish to emphasize some main differences between \MadOnia\ and \HELACOnia. \HELACOnia\ is based on recursion relations to calculate helicity amplitudes, while \MadOnia\ uses the traditional Feynman diagrams. Moreover, \HELACOnia\ is designed to deal with processes containing one or more heavy quarkonium up to P-wave Fock states, while the number of states of heavy quarkonia is restricted to one in \MadOnia.

The aim of this writeup is to introduce a 2.0 version of \HELACOnia, where many new and useful features are included, which are motivated by the practical phenomenological simulations and the user experience, e.g. interfacing with parton shower Monte Carlo programs. In section \ref{sec:meth}, we will describe the methodology, the related algorithms and the new features in \HELACOnia\ 2.0. Then, we will show how to use the program in section \ref{sec:usage}. Several examples are given in section \ref{sec:exam}. In section \ref{sec:con}, we draw our conclusions. Some useful information are given in the appendices. The program structure is sketched in appendix \ref{sec:program}. A summary of the new particle symbols in \HELACOnia\ 2.0 is tabulated in appendix \ref{app:symbol} and a few of useful new parameters are introduced in appendix \ref{app:param}. Finally, the addon codes in \HELACOnia\ 2.0 are introduced in appendix \ref{sec:addon}.

\section{Methodology, algorithm and new features\label{sec:meth}}
\subsection{Heavy-quarkonium-amplitude computation with recursion relations}

As it was introduced in our previous document~\cite{Shao:2012iz},\HELACOnia\ is based on the off-shell recursion relation~\cite{Berends:1987me}. \HELACOnia\ is based on a public package \HELAC~\cite{Kanaki:2000ey,Papadopoulos:2006mh,Cafarella:2007pc}, which
is based on the Dyson-Schwinger equations~\cite{Dyson:1949ha,Schwinger:1951ex,Schwinger:1951hq} to calculate the helicity amplitude in the SM at parton level. In this section, we will first briefly recall how to calculate a helicity amplitude for a general process with $n$ external legs. We denote the momenta of these
external legs as $p_1,p_2,\ldots,p_n$, and their quantum
numbers, such as color and helicity, are symbolized as
$\alpha_1,\alpha_2,\ldots,\alpha_n$. Any $k$ external legs can form an off-shell current as $\mathcal{J}(\{p_{i_1},\ldots,p_{i_k}\};\{\alpha_{i_1},\ldots,\alpha_{i_k}\})$.We can assign each current $\mathcal{J}$ a number $l$, which is called ``level". It is defined
as the number of external legs involved in the current $\mathcal{J}$, i.e. the
``level" of
$\mathcal{J}(\{p_{i_1},\ldots,p_{i_k}\};\{\alpha_{i_1},\ldots,\alpha_{i_k}\})$
is $k$. The construction of the higher ``level" currents is from the lower ``level" ones in a recursion relation, where the starting point of the recursion
relation is external legs and its end point is to obtain the ``level" $n$ current. The advantages by working in this way is that one is able to avoid computing identical subgraphs
contributing to different Feynman diagrams more than once. The
summation of all subgraphs contributing to a specific
current reduces the total number of objects that should be used in
the next recursion procedure. 

In \HELACOnia, we calculate the heavy-quarkonium amplitude in the framework of NRQCD factorization.   With this formalism, the cross section for a heavy quarkonium $\mathcal{Q}$ production can be factorized into the
perturbative short-distance components and the non-perturbative
long-distance matrix elements (LDMEs). For example, at a proton-proton collider, the cross section can be written as
\bq\sigma(pp\to\mathcal{Q}+X)=\sum_{i,j,n}{\int{\rm{d}x_1\rm{d}x_2{\textit f}_{i/p}(x_1){\textit f}_{j/p}(x_2)
\hat{\sigma}(ij\to
Q\bar{Q}[n]+X)\langle\mathcal{O}^{\mathcal{Q}}_n\rangle}},\eq
where
$\textit{f}_{i/p}$ and ${\textit f}_{j/p}$ are the parton
distribution functions (PDFs),$\hat{\sigma}(ij\to Q\bar{Q}[n]+X)$ is the
short distance coefficient to produce a heavy quark pair
$Q\bar{Q}$ in the specific quantum state $n$. Following the usual notation, the Fock states $n$ can be written in the spectroscopic form $n={\bigl.^{2S+1}\hspace{-1mm}L^{[c]}_J}$, where $S,L,J$ identify
the spin, orbital momentum, total angular momentum states
respectivley, and $c=1,8$ means that the intermediate state
$Q\bar{Q}$ can be in a color-singlet or a color-octet state. The LDMEs are denoted as $\langle\mathcal{O}^{\mathcal{Q}}_n\rangle$. Their physical interpretation is a probability\footnote{Rigorously speaking, the LDMEs are not physical, while only the (differential) cross section is. They are much like the PDFs and fragmentation functions (FFs).} for a heavy quark pair in the Fock state $n$ to evolve into a quarkonium. The power counting
rules in NRQCD yield to the fact that for any quarkonium, there should be only a limited number of
Fock states contributing to a specific order of $v$, where $v$ is the relative velocity of the heavy quark pair. The projection method is used to project a heavy-flavor quark pair onto a specific Fock state. The color-singlet projector is $\frac{\delta_{ij}}{N_c}$, while the color-octet projector is $\sqrt{2}\lambda^a_{ij}$, where $\lambda^a$ is the Gell-Mann matrix and it will be projected further onto a color-flow basis~\cite{'tHooft:1973jz,Kanaki:2000ms,Maltoni:2002mq}. The spin projectors~\cite{Berger:1980ni} are in the form of
\bq-\frac{1}{2\sqrt{2}(E+m_{Q})}\bar{v}(p_2,\lambda_2)\Gamma_S\frac{\slashed{P}+2E}{2E}u(p_1,\lambda_1),\label{eq:spinpro}\eq
where $m_{Q}$ is the mass of the heavy quark, $p_1,p_2$ and
$\lambda_1,\lambda_2$ are the momenta and helicity of the heavy quarks
respectively. The total momentum of the heavy quark pair is $P^{\mu}=p^{\mu}_1+p^{\mu}_2$ and $E=\frac{\sqrt{P^2}}{2}$. $\Gamma_S$ is
$\gamma_5$ for the spin singlet state $S=0$, and it is
$\epsilon_{\mu}^{\lambda_s}\gamma^{\mu}$ for the spin triplet state $S=1$ with $\lambda_s=\pm,0$. In order to construct the ``level" 1 current for the heavy quarkonium, we cut the
fermion chain at the place of $\slashed{P}+2E$ in the projector shown in
Eq.(\ref{eq:spinpro}). Then, the 
new ``level" $l=1$ current for $Q$ as
$\frac{1}{m_Q}\bar{u}(P,\lambda^{\prime})(\slashed{p}_1+m_Q)$ and
for $\bar{Q}$ as
$-\frac{1}{8\sqrt{2}m_Q}(\slashed{p}_2-m_Q)u(P,\lambda^{\prime})$. \HELACOnia\ is also designed to be able to handle P-wave states. In \HELACOnia, we introduced numerical stable P-wave currents, which have already been  discussed in Ref.~\cite{Shao:2012iz}.

\subsection{Angular distributions of heavy-quarkonium decays \label{subsec:angdis}}

Angular distributions of heavy-quarkonium decays have attracted a lot of attention in the past few years.  Theorists are interested in the polarization observables of heavy-quarkonium production because they might provide a ``smoking gun" to discriminate the various heavy-quarkonium-production mechanisms. The understanding of heavy-quarkonium polarization is also crucial for the simulations in the experimental analyses, e.g the detector acceptance for lepton pairs  from the decay of $J^{\rm PC}=1^{--}$ heavy quarkonium strongly depends on its polarization or on the angular distribution of its decay products. The crue implementation of the polarization in the simulation of heavy-quarkonium production usually leads to one of the largest systematic uncertainties on the measurements. However, the current available Monte Carlo programs like \Pythia\ are usually very limited in the available decay processes of heavy quarkonium and/or assume the unpolarized pattern. Hence, it is our motivation to implement some frequently used decay processes of heavy quarkonium with the polarization pattern.

For the simple decay processes, like $J/\psi\to\ell^+\ell^-$, we only have to follow the polarization of the mother particle and the implementation of the angular distribution in each spin eigenstate is straightforward. However, for a general decay process, e.g. $\chi_{c,J}\to J/\psi+\gamma, J=1,2$, the algorithms for generating the angular distributions of the decay products is following: 
\begin{enumerate}
\item Considering the helicity amplitude for the decay process is $\mathcal{A}({\bf x})$, where ${\bf x}$ is the set of variables to characterize the kinematics of the decay process.
\item The maximal weight of $|\mathcal{A}({\bf x})|^2$ is $W_{\rm max}$.
\item Randomly generate a phase space point ${\bf x}$.
\item Uniformly generate a random number $r\in [0,1]$. If $|\mathcal{A}({\bf x})|^2>r\times W_{\rm max}$, the event corresponding to ${\bf x}$ is retained. Otherwise, go to the former step.
\end{enumerate}
All of the available hard-coded decay processes in \HELACOnia\ 2.0 can be found in \textbf{decay}/\textbf{decay\_list.txt}. \HELACOnia\ 2.0 also supports the cascade decays. For instance, a generated $\chi_{c,J}$ meson can be decayed into  a $J/\psi$ and a photon $\gamma$ first and then the decay product $J/\psi$ can be further decayed into a lepton pair. However, the first decay process may depend on the true masses of $\chi_{c,J}$ and $J/\psi$. The corresponding input values can be changed by the user in \textbf{input}/\textbf{decay\_param\_user.inp}.

Such a module is flexible and can be extended to other decay processes. Especially, we are planning to include \EvtGen~\cite{Lange:2001uf,Ryd:2005zz} for B-physics studies in \HELACOnia.

\subsection{Interface to parton shower Monte Carlo programs}

Parton shower Monte Carlo programs are widely used in numerical simulations for the collider enviroment.

\subsubsection{\HELACOnia+\Pythia8\label{subsec:py8}}

\Pythia\ is a general purpose Monte Carlo program. It provides the QCD and QED parton shower as well as the hadronization. Experiments performed on high-energy colliders rely heavily on it. Hence, the interface between \HELACOnia\ and \Pythia\ would surely extend the applications of the program. \HELACOnia\ 2.0 has indeed been sucessfully interfaced with \Pythia8\footnote{Currently, it only works for \Pythia8.1~\cite{Sjostrand:2007gs} but not for \Pythia8.2~\cite{Sjostrand:2014zea}.}, which is written in \Cpp. Its usage in \HELACOnia\ requires the user to pre-install \HepMC~\cite{Dobbs:2001ck} and \Pythia8~\cite{Sjostrand:2007gs}. Inheriting to its processor \Pythia6~\cite{Sjostrand:2006za}, \Pythia8 provides an interface to the external hard matrix element/event generators via Les Houches Event files~\cite{Alwall:2006yp} according to Les Houches accord format~\cite{Boos:2001cv}. In \HELACOnia\ 2.0, several files in \Cpp\ are written to use the generated Les Houches files and to shower and to hadronize the unweighted events with \Pythia8 on the fly. The default output is \HepMC\ event file after passing through \Pythia8. However, such a format is usually inefficient to store events since it might result in a huge \HepMC\ file from a relative large Les Houches file (say one million events). Two alternative options are provided. One is to output \TopDrawer\ format plots with \Hbook.\footnote{We used a simplified version of \Hbook\ written by M. Mangano.} However, such option requires the user to define all of the observables and the histograms in \Fortran\ 90 before calling \Pythia. Useful analysis tools, like \FastJet~\cite{Cacciari:2011ma} (or \FJCore ) and \HEPTopTagger~\cite{Plehn:2009rk,Plehn:2010st,Plehn:2011sj}, can also be linked to fill the histograms. Some examples are given in \textbf{analysis}/\textbf{PYTHIA8}. Another option is using the software \Root, which however requires the user to pre-install \Root. Events after showering and hadronization will be filled into \Root\ tree, and a pre-defined \Cpp\ \Root\ script is necessary. We also give some examples in the subdirectory \textbf{analysis}/\textbf{PYTHIA8}. We will described its detailed usage in section \ref{sec:usage}.

In order to read the events record in a \HepMC\ file, we also provide some useful tools for converting it to a \TopDrawer\ file or a \Root\ tree file in \HELACOnia\ 2.0. Their exectuable files are \textbf{HepMC2Plot} and \textbf{HepMC2Root} in the directory \textbf{bin}.

In principle, such a methodology can be applied to the interfaces to other parton shower Monte Carlo programs as long as it can recognize the Les Houches event files. Although the current version of \HELACOnia\ is still not interfaceable with other general-purpose parton shower Monte Carlo programs automatically, it is in our to-do list to write the similar interfaces for \Pythia6~\cite{Sjostrand:2006za},\Herwig6~\cite{Corcella:2000bw,Corcella:2002jc} and \Herwig++~\cite{Bahr:2008pv,Bellm:2013lba}.

\subsubsection{\HELACOnia+\QEDPS}

\QEDPS\ is a program for the photon showering from the initial $e^{\pm}$ in electron-positron collisions~\cite{Fujimoto:1993qh,Fujimoto:1993ge,Munehisa:1995si,Kato:1988ii}, which is based on the fact that the structure function of an electron $D_{e^{\pm}}(x,Q^2)$ should obey the Altarelli-Parisi equation
\bqa
\frac{dD_{e^{\pm}}(x,Q^2)}{d\log{Q^2}}=\frac{\alpha}{2\pi}\int^1_{x}{\frac{dy}{y}P_{+}\left(\frac{x}{y}\right)D_{e^{\pm}}(y,Q^2)},
\eqa
where $x$ is the longitudinal fraction, $Q^2$ is the virtuality and $P_{+}(x)$ is the Altarelli-Parisi splitting function. In the leading logarithm approximation, one can solve it to be
\bqa
D_{e^{\pm}}(x,Q^2)=\Pi(Q^2,Q_s^2)D_{e^{\pm}}(x,Q_s^2)+\frac{\alpha}{2\pi}\int^{Q^2}_{Q_s^2}{\frac{ds}{s}\Pi(Q^2,s)\int^{1-\epsilon}_{x}{\frac{dy}{y}P(y)D_{e^{\pm}}\left(\frac{x}{y},s\right)}},
\eqa
where $Q_s^2\sim m_e^2$ and $\Pi$ is the Sudakov factor
\bqa
\Pi(Q^2,Q_0^2)=\exp\left(-\frac{\alpha}{2\pi}\int^{Q^2}_{Q_0^2}{\frac{ds}{s}\int_0^{1-\epsilon}{dx P(x)}}\right).
\eqa
As it is well known, the Sudakov factor has the interpretation of the probability of the electron evolving from scale $Q_0^2$ to $Q^2$ without emitting any hard photon. We review the algorithm of the photon shower in \QEDPS~\cite{Fujimoto:1993qh,Fujimoto:1993ge,Munehisa:1995si,Kato:1988ii}:
\begin{enumerate}
\item Initially, set $x=1$, where $x$ is the fraction of the light-cone momentum of the electron/positron.
\item Generate a random number $r$. If $r$ is smaller than $\Pi(Q^2,Q_s^2)$, the evolution stops. Otherwise, try to find the next virtuality $Q^2$ with $r=\Pi(Q^2,Q_s^2)$. At $Q^2$, a branching $e^{\pm}\to e^{\pm}(Q^2)\gamma$ happens.
\item According to the probability of the splitting function $P(y)$ between 0 and $1-\epsilon$, try to determine the value of $y$. Replace the original $x$ to be $xy$. Iterate the step 2 until the evolution stops.
\end{enumerate}
Hence, \QEDPS\ provides a leading logarithm approximation to the initial state radition in electron-positron collisions.

\HELACOnia\ 2.0 provides an interface to \QEDPS\ when calculating an eletron-positron annihilation process. Thanks to the release of the source code, \QEDPS\ is self contained and the user does not need to install it by himself/herself. An input parameter {\tt emep\_ISR\_shower} is provided in \textbf{input}/\textbf{user.inp} or \textbf{input}/\textbf{default.inp} to determine whether using \QEDPS\ to perform a photon shower (see its description in appendix \ref{app:param}). If one wants to generate \TopDrawer, \Gnuplot\ or \Root\ plot files, one should also edit the \Fortran\ subroutine {\tt plot\_fill\_QEDPS} before compilation. We have performed an application of \HELACOnia+\QEDPS\ to $J/\psi$ inclusive production at B factories in Ref.~\cite{Shao:2014rwa}.

\subsection{Estimating the scale and PDF uncertainties}

Varying the renormalization scale $\mu_R$ and the factorization scale $\mu_F$ is often thought to be a standard way to estimate the theoretical uncertainty in perturbative computations due to the missing higher-order contributions. Although such an argument can be applied to the scattering or decay processes in general as long as its validation of the perturbative description, there are indeed several cases which we already encountered where it is not applicable, such as where one encounters large coefficients correction (from large logarithms, large $\pi^2$, or large color factors)
or new channels (e.g. new initial states, new phase space region, or new fragmentation topology). Unfortunately, the later case frequently happens in heavy-quarkonium-production processes. Because new fragmentation topologies appear only at higher orders in perturbative calculations, it is usually necessary to consider the contributions from the higher-multiplicity processes accompanying with more partons. Some examples indeed already can be seen in the single-quarkonium~\cite{Campbell:2007ws,Artoisenet:2008fc} and double-quarkonium~\cite{Lansberg:2013qka} production processes. Nevertheless, after taking into account all of the important topologies, the scale dependence is sufficiently reasonable to estimate the higher-order corrections. At tree level, the renormalization-scale dependence is only in the renormalization running of $\alpha_s(\mu_R)$, while the factorization-scale dependence is in the Dokshitzer-Gribov-Lipatov-Altarelli-Parisi (DGLAP) evolution~\cite{Altarelli:1977zs,Dokshitzer:1977sg,Gribov:1972ri} of the PDFs. Hence, it is straightfoward that the estimation of the scale uncertainties are irrevelant to the most time-consuming matrix-element calculations as long as one knows the initial states and the perturbative orders. From the technical point of view, such estimation can be zero CPU cost.\footnote{In a general case, the explicit renormalization and factorization scales dependence is also known at next-to-leading order (NLO)~\cite{Frederix:2011ss}.} In calculating a physical observable or filling a histogram, one just multiplies a weight 
\bqa
w_{\rm scale}(\mu_R,\mu_F;\mu_R^0,\mu_F^0)\equiv \frac{f_{1}(\mu_F,x_1)f_{2}(\mu_F,x_2)}{f_{1}(\mu_F^0,x_1)f_{2}(\mu_F^0,x_2)}\left(\frac{\alpha_s(\mu_R)}{\alpha_s(\mu_R^0)}\right)^b
\eqa
to the central value in each phase space point or each event, where $f_i$ is the PDF, $x_i$ is the Bjorken fraction, $\mu_{R,F}$ ($\mu_{R,F}^0$) is the new (central) renormalization and factorization scales, $b$ is the power of $\alpha_s$ in the squared amplitude. Such reweighting procedure has been widely used in other programs, such as \MadGraph~\cite{Alwall:2007st,Alwall:2011uj,Alwall:2014hca}.

Another important source of theoretical uncertainty that can be obtained from the reweighting method is the PDF uncertainty, which does not reflect the uncertainty in the hard matrix element but rather the uncertainty in the extraction of the PDF. It is known that the PDF uncertainty stems from at least three sources: the uncertainties in the input (experimental) data, the accuracy of the perturbative calculation, and the method to extract the PDF. Most of the modern PDF sets provide a way to estimate the impact of such uncertainty to the theoretical calculations. For example, the global-fit PDF MSTW 2008 NLO~\cite{Martin:2009iq} provides $40$ error PDFs to quantify its uncertainty. Instead of reevaluating the matrix element with new PDF, one is able to evaluate the PDF uncertainty by multiplying a weight
\bqa
w_{\rm pdf}(f_1^{\prime},f_2^{\prime};f_1,f_2)\equiv \frac{f^{\prime}_{1}(\mu_F,x_1)f^{\prime}_{2}(\mu_F,x_2)}{f_{1}(\mu_F,x_1)f_{2}(\mu_F,x_2)},
\eqa
where $f_i^{\prime}$ is an error PDF and $f_i$ is the central PDF. Such a procedure is exact in a parton-level calculation. However, it should be understood as an approximation when incorporating with a parton shower Monte Carlo program, since the backward evolution of the initial state partons in Monte Carlo indeed contains an implicit dependence on the chosen PDF, i.e. the central one in a Les Houches event file.

Although such a procedure is more trivial at tree level than that at NLO~\cite{Frederix:2011ss}, for completeness, we would like to emphasize that \HELACOnia\ 2.0 provides a functionality to estimate scale and PDF uncertainties with such a reweighting method. However, because of the recursion relations, it is non-trivial to separate different coupling orders without degrading its speed advantage. Hence, we want to remind the user that it would be wrong in evaluating renormalization scale dependence if the amplitudes in different orders will contribute to a partonic level process.

\subsection{A summary of new features}

We give a summary before closing this section. \HELACOnia\ 2.0 has been improved much compared to the first released version of \HELACOnia~\cite{Shao:2012iz}. The main changes include:
\begin{enumerate}
\item Two completely independent generators based on \PHEGAS~\cite{Papadopoulos:2000tt} and VEGAS~\cite{Lepage:1977sw} are implemented. Both of them can generate unweighted events for $2 \to n$ processes when $n\ge2$ at $pp$,$p\bar{p}$ and $e^-e^+$ collisions. For $2\to 1$ processes at hadron colliders, only VEGAS is available.
\item Additional internal PDFs are included. The program can also be interfaced with LHAPDF~\cite{Whalley:2005nh}.
\item Analysis plots are done on the fly. Differential distributions can be plotted at the end of the {\it computation phase}.
\item The laboratory frame is not restricted to the center-of-mass frame of the initial collision anymore. The fixed-target mode was also added.
\item An interface from \QEDPS\ to \HELACOnia\ is done. One can include the QED photon showering effects from initial $e^{\pm}$ beams.
\item An interface from \Pythia\ 8 to \HELACOnia\ is done. It is able to use the major functionality in \Pythia\ 8.
\item Reweighting method is used to estimate the renormalization/factorization scale and PDF uncertainties.
\item Several spin-entangled decay processes are implemented to take into account the polarization effects.
\item A user-friendly \Python\ script is provided for user to use the program. It also allows us to calculate the cross sections of several subprocesses with multiply CPUs, such as on a multicore computer or on a cluster.
\end{enumerate}

For item 1, in the previous version, the unweighted events can only be generated by \PHEGAS~\cite{Papadopoulos:2000tt}. $2\to 1$ processes are not handled in this case. This improvement allows us to lift several restrictions. Concerning item 2, beforehand, only CTEQ6~\cite{Pumplin:2002vw} was available. It paves the way to application in nucleus collisions and to estimate PDF uncertainties. With the help of the improvement presented in item $4$, we are able to apply \HELACOnia\ to more experiments like fixed-target experiments~\cite{Brodsky:2012vg}. Item $5$ allows us to consider initial radiation effects at $e^-e^+$ collision, which might not be small in several important processes,such as $e^-e^+\to \jpsi+gg$~\cite{Shao:2014rwa}, while the improvement of item $6$ extends the usage of \HELACOnia\ to 
one of the most widely used multipurpose event generator \Pythia\ 8. For item $7$, it will be very useful to estimate the renormalization/factorization scale and PDF uncertainties without the extra cost of recalculating cross sections. The improvement of item $8$ is quite  useful for practical simulations, and that of item $9$ improves the user experience on using the program.

\section{How to use the program\label{sec:usage}}

In this section, we will first give a brief introduction on how to perform a phenomenological analysis in a basic way. If one is only interested in using the program, one can follow the instruction in this section and ignore the remaining context. 

\subsection{Standalone}

We first introduce how to use \HELACOnia\ 2.0 in a standalone way. Before running the code, one should specify the configurations via the configuration file \textbf{ho\_configuration.txt} in the \textbf{input} subdirectory, which is described in section \ref{sec:program}. For example, if one wants to use LHAPDF~\cite{Whalley:2005nh}, one should assign the correct path to the parameter {\tt lhapdf} in the file \textbf{ho\_configuration.txt}. Useful comments are also given in the configuration file. If one wants to output plot(s) on the fly, one should also edit the user plot file \textbf{plot\_user.f90} in the subsubdirectory \textbf{analysis}/\textbf{user}. One can follow some example files to write his/her own plot file. After the above preparations, one can set the configuration and make the files via the command line
\cCode{}
> ./config
\end{lstlisting}
This procedure should only be done once. Afterward, the program is ready for running.

In \HELACOnia\ 2.0, one can use two modes to perform a computation of a cross section. We still keep the initial way to run the program directly via exectuable file \textbf{Helac-Onia}. To use this way, one should follow the following lines:
\begin{enumerate}
\item Specify input parameters in \textbf{input}/\textbf{user.inp} following the format in \textbf{input}/\textbf{default.inp}. Some examples are given in \textbf{input}/\textbf{demo}.
\item Provide the process information in \textbf{input}/\textbf{process.inp} as well as the decay information in \textbf{input}/\textbf{decay\_user.inp}.
\item If one wants to define one's own dynamical renormalization and/or factorization scale, one should edit it in \textbf{src}/\textbf{setscale.f90} before compiling. Four default scales are defined, i.e. the fixed scale, the transverse mass $m_{T,1}=\sqrt{m_1^2+P_{T,1}^2}$ of the first final state, $\mu_0=\sqrt{(\sum_{f\in {\rm final~states}}{m_f})^2+P_{T,1}^2}$ and $\mu_0=\frac{1}{2}\sum_{f\in {\rm final~states}}{\sqrt{m_f^2+P_{T,f}^2}}$.
\item Run the program with the command line
\cCode{}
> ./Helac-Onia
\end{lstlisting}
or
\cCode{}
> ./bin/Helac-Onia
\end{lstlisting}
The final results will be collected in the output directory.
\end{enumerate}

A second way to run \HELACOnia\ 2.0 is by using the \Python\ scripts, which is more user-friendly and hence strongly recommended. It provides the opporturnity to avoid mixing the working directory and the \HELACOnia\ directory. If one wants to use the majority of the new features in the program, one has to run the program with the \Python\ scripts. The basic way of using it after the above first three items is to run the program with the script
\cCode{}
> /PATH/TO/HELAC-Onia/ho_cluster
\end{lstlisting}
where /PATH/TO/HELAC-Onia is a path to the \HELACOnia\ directory relative to your working directory. Then one will see a prompt starting with ``$\rm{HO}\hspace{-1mm}>$". There are two phases to compute a cross section for a process, i.e. {\it generate the process} and {\it run the program}. In the first phase, one should define one process or several subprocesses. For example, if one wants to calculate $J/\psi$ pair production, the syntax should be
\cCode{}
HO> generate g g > cc~(3S11) cc~(3S11)
\end{lstlisting}
where the symbol $g$ represents the initial gluon and $cc\sim(3S11)$ means a pair of charm (anti-)quark $c$ and $\bar{c}$ in $\ss$ configuration, which can be found in appendix \ref{app:symbol}. One can also calculate the cross sections for several subprocesses simultaneously. For instance, if one wants to go beyond the leading-order computation of $J/\psi$ pair production, he/she can type the following command lines
\cCode{}
HO> generate g g > cc~(3S11) cc~(3S11)
HO> generate g g > cc~(3S11) cc~(3S11) g
HO> generate u g > cc~(3S11) cc~(3S11) u
HO> generate g u > cc~(3S11) cc~(3S11) u
\end{lstlisting}
It will use multiple cores to calculate the cross sections on the cluster or on a multicore machine. One can also run the addon processes, where the available addon processes and the corresponding numbers are listed in \textbf{addon}/\textbf{addon\_process.txt}. For instance, if one wants to calculate the double parton scattering (DPS) for $J/\psi$ pair production, one can generate the process via
\cCode{}
HO> generate addon 1
\end{lstlisting}
where the keyword {\it addon} should be specified after {\it generate} and the number for this process is 1 as be seen in \textbf{addon}/\textbf{addon\_process.txt}. Before launching the jobs for numerical calculations, one can also change the parameters in \textbf{input}/\textbf{user.inp} via the interactive command syntax
\cCode{}
HO> set <parameter_name> = <value>
\end{lstlisting}
One example is to take the maximum Monte Carlo integration number to be $10000$. Then one just simply types
\cCode{}
HO> set nmc = 10000
\end{lstlisting}
If one wants to take VEGAS as the Monte Carlo integration program, please uses
\cCode{}
HO> set gener = 3
\end{lstlisting}
Another useful feature is to define the decay process(es) via \Python\ scripts. The syntax is
\cCode{}
HO> decay <process> @ <branching ratio>
\end{lstlisting}
The command lines
\cCode{}
HO> decay cc~(3S11) > m+ m- @ 0.06d0
HO> decay w+ > m+ vm @ 1d0
\end{lstlisting}
means that the $\ss$ charmonium in the final states will decay to a muon pair with the branching ratio $6\%$ and the $W^+$ boson will perform leptonic decay to a muon and a neutrino with $100\%$ probability. After all, one just submits the job via command
\cCode{}
HO> launch
\end{lstlisting}
and waits for the final results to be collected in a new created subsubdirectory \textbf{PROC\_HO\_i}/\textbf{results} in the working directory, where \textbf{i} is a number to be assigned uniquely.

In the new way, we take a similar fashion of the widely used program \MG5aMC, and we hope that it would become a standard in the future, or at least it will be much easier for the users who are already familiar with \MG5aMC\ to learn to use this program.

\subsection{\HELACOnia+\Pythia8}

Let us start to consider the case of using \HELACOnia+\Pythia8.\footnote{It is more or less trivial to use \HELACOnia+QEDPS by setting the flag {\tt emep\_ISR\_shower} to be 1 in \textbf{input}/\textbf{user.inp} as described in appendix \ref{app:param}.} One has to generate the Les Houches file for unweighted events in \textbf{PROC\_HO\_i}/\textbf{P0\_calc\_j}/\textbf{output} first before calling \Pythia. This can be achieved by setting the flag {\tt unwgt} to be T before launching the program. 

The program will be able to call \Pythia8 if one sets correct path for \Pythia8 and \HepMC\ in the configuration file
\textbf{ho\_configuration.txt} and specifies the shower parameters\footnote{Especially, one should be aware of the parameter {\tt ANALYSE} in \textbf{shower\_card\_user.inp}. It determines the output mode as described in section \ref{subsec:py8}. If {\tt ANALYSE} is empty, it will output a \HepMC\ event file from \Pythia. One can also take {\tt ANALYSE} to be a \Fortran 90 or \Cpp\ plot file in the other two output modes in section \ref{subsec:py8}. For example, if {\tt ANALYSE} = {\tt plot\_py8\_pp\_tj} and there exist a file {\tt plot\_py8\_pp\_tj.f90} in \textbf{analysis}/\textbf{PYTHIA8} in the \HELACOnia\ root directory, it will output a \TopDrawer\ file from \Pythia8. If the extension of the file is {\tt .cc} in \textbf{analysis}/\textbf{PYTHIA8}, it will output a \Root\ file instead.} in the file \textbf{shower\_card\_user.inp} before using the \textbf{ho\_cluster} script. The corresponding parameter setup for calling \Pythia8 in \textbf{user.inp} is 
\cCode{}
HO> set parton_shower = 1
\end{lstlisting}

All of the above shower related setup can be done before or after generating Les Houches event files. In the former case, the \Pythia8 will be called with the command {\tt launch} directly, while in the later case we also provide a new command {\tt shower} with the syntax
\cCode{}
HO> shower <working path>
\end{lstlisting}
where the working path is usually \textbf{PROC\_HO\_i}. The final output from \Pythia8 will be collected in the directory \textbf{PROC\_HO\_i}/\textbf{P0\_calc\_j}/\textbf{shower}/\textbf{HO\_PYTHIA8\_k}, where \textbf{i},\textbf{j} and \textbf{k} are integers starting from 0. 

We will present some technical details for using \Pythia8 in \HELACOnia\ 2.0 directly, since the user might have encountering problems in compiling the \Pythia8 related code with wrong setup. It will also be useful for extensive usage of \Pythia8 in \HELACOnia. After running the program with {\tt launch} or {\tt shower} commands with the correct shower-related setup, a directory \textbf{HO\_PYTHIA8\_k} will be created in \textbf{PROC\_HO\_i}/\textbf{P0\_calc\_j}/\textbf{shower}. If the program compiles successfully, an exectuable file \textbf{Pythia8.exe} will be generated. One can also change the \Pythia8 setup via the file \textbf{Pythia8\_lhe.cmnd} in the same directory. A direct using \Pythia8 is possible by simply typing
\cCode{}
> ./Pythia8.exe
\end{lstlisting}
If there does not exist the exectuable file \textbf{Pythia8.exe}, it means there is a problem in the compilation. Some useful information can be found in \textbf{shower.log} to solve the problems.


%


\section{Examples\label{sec:exam}}

\subsection{NNLO$^{\star}$ level $J/\psi$ and $\psi(2S)$ hadroproduction}

$\psi$ and $\Upsilon$ production at hadron colliders have challenged our understanding of heavy-quarkonium mechanism for decades~\cite{Abe:1992ww,Abe:1997jz,Abe:1997yz}. Since then, the heavy-quarkonium-production data have been removed in the global fit of extracting PDF. Large QCD corrections were found in heavy-quarkonium production due to new $p_T$-enhanced fragmentating Feynman diagrams at higher orders~\cite{Campbell:2007ws}. Hence, it was suggested to look at how partial next-to-next-to-leading order (NNLO) QCD correction impacts the differential cross sections of $\psi$ and $\Upsilon$ production~\cite{Artoisenet:2008fc}, which was called NNLO$^{\star}$. Besides of the complete NLO result~\cite{Campbell:2007ws}, it requires to calculate the $\mathcal{O}(\alpha_s^5)$ tree-level $2\to 4$ process $pp\to \psi$+3-jets. \MadOnia~\cite{Artoisenet:2007qm} was able to perform a first numerical computation with such complex. It is a good process to show the robustness of \HELACOnia\ and to compare \MadOnia.

The calculation of $\mathcal{O}(\alpha_s^5)$ process $p p \rightarrow \psi$+3-jets in the color-singlet mechanism (CSM) consists the following $13$ independent subprocesses:
\cCode{}
HO> generate g g > cc~(3S11) g g g
HO> generate g g > cc~(3S11) u u~ g
HO> generate u g > cc~(3S11) g g u
HO> generate g u > cc~(3S11) g g u
HO> generate u g > cc~(3S11) u u~ u
HO> generate g u > cc~(3S11) u u~ u
HO> generate u g > cc~(3S11) d d~ u
HO> generate g u > cc~(3S11) d d~ u
HO> generate u u~ > cc~(3S11) g g g
HO> generate u u~ > cc~(3S11) u u~ g
HO> generate u u~ > cc~(3S11) d d~ g
HO> generate u u > cc~(3S11) u u g
HO> generate u d > cc~(3S11) u d g
\end{lstlisting}
In the above command lines, we only include up and down quarks. The other quark-initial-state contribution can be included by
\cCode{}
HO> set quarksumQ = T
HO> set iqnum = 3
\end{lstlisting}
The truth of the flag {\tt quarksumQ} makes sure we will include the initial (anti)quark PDFs and {\tt iqnum}=3 means we are working in 3-light-quark-flavor scheme. Moreover, one can type
\cCode{}
HO> set combine_factors = 1. 3. 1. 1. 1. 1. 2. 2. 1. 1. 2. 1. 1.
\end{lstlisting}
to explicitly multiply a combination factor for each subprocess. For example, in the subprocess $gg\rightarrow c\bar{c}[\ss]+u \bar{u} g$, we take a factor $3$ to account for $gg\rightarrow c\bar{c}[\ss]+q \bar{q} g$ with $q=u,d,s$, because they share the same matrix element and PDF. The detailed correspondances are shown in Tab.\ref{tab:nnlosubproc}. In order to avoid infrared divergence in NNLO$^{\star}$ calculations, a special cutoff $s_{ij}^{\rm min}$~\cite{Artoisenet:2008fc} should be applied to the invariant mass squared of any massless parton pair, i.e. $(p_i+p_j)^2\ge s_{ij}^{\rm min}$. In this case, we set 
\cCode{}
HO> set minmqqp = 3d0
\end{lstlisting}
for any final state massless parton pair $\sqrt{(p_i+p_j)^2}\ge 3.0$, and set
\cCode{}
HO> set minmqbeam = 3d0
\end{lstlisting}
for one final state massless parton and a initial state parton $\sqrt{-(k_{1,2}-p_i)^2}\ge 3.0$.

\begin{table}
\begin{center}
\begin{tabular}{{c|}*1 c} 
\hline\hline
Syntax & Subprocess \\\hline
\chCode{}
g g > cc~(3S11) g g g
\end{lstlisting}& $gg\rightarrow c\bar{c}[\ss]+ggg$ \\\hline
\chCode{}
g g > cc~(3S11) u u~ g
\end{lstlisting}&  $gg\rightarrow c\bar{c}[\ss]+q\bar{q}g$\\
~ & with $q=u,d,s$  \\\hline
\chCode{}
u g > cc~(3S11) g g u
\end{lstlisting} &  $qg\rightarrow c\bar{c}[\ss]+ggq$ \\
~ & with $q=u,\bar{u},d,\bar{d},s,\bar{s}$ \\\hline
\chCode{}
g u > cc~(3S11) g g u
\end{lstlisting} &  $gq\rightarrow c\bar{c}[\ss]+ggq$\\
~ & with $q=u,\bar{u},d,\bar{d},s,\bar{s}$ \\\hline
\chCode{}
u g > cc~(3S11) u u~ u
\end{lstlisting}  &  $q g \rightarrow c\bar{c}[\ss]+q\bar{q}q$\\
~ & with $q=u,\bar{u},d,\bar{d},s,\bar{s}$ \\\hline
\chCode{}
g u > cc~(3S11) u u~ u
\end{lstlisting} & $gq \rightarrow c\bar{c}[\ss]+q\bar{q}q$\\
~ & with $q=u,\bar{u},d,\bar{d},s,\bar{s}$ \\\hline
\chCode{}
u g > cc~(3S11) d d~ u
\end{lstlisting} &  $qg \rightarrow c\bar{c}[\ss]+q^{\prime}\bar{q^{\prime}}q$\\
~ & with $q=u,\bar{u},d,\bar{d},s,\bar{s}$\\
~ & and $q^{\prime}=u,d,s$\\
~ & and $q,q^{\prime}$ not in the same flavor\\\hline
\chCode{}
g u > cc~(3S11) d d~ u
\end{lstlisting} & $gq \rightarrow c\bar{c}[\ss]+q^{\prime}\bar{q^{\prime}}q$\\
~ & with $q=u,\bar{u},d,\bar{d},s,\bar{s}$ \\
~ & and $q^{\prime}=u,d,s$ \\
~ & and $q,q^{\prime}$ not in the same flavor \\\hline
\chCode{}
u u~ > cc~(3S11) g g g
\end{lstlisting} & $q\bar{q}\rightarrow c\bar{c}[\ss]+ggg$ \\
~ & with $q=u,\bar{u},d,\bar{d},s,\bar{s}$ \\\hline
\chCode{}
u u~ > cc~(3S11) u u~ g
\end{lstlisting} & $q\bar{q}\rightarrow c\bar{c}[\ss]+q\bar{q}g$ \\
~ & with $q=u,\bar{u},d,\bar{d},s,\bar{s}$\\\hline
\chCode{}
u u~ > cc~(3S11) d d~ g
\end{lstlisting} & $q\bar{q}\rightarrow c\bar{c}[\ss]+q^{\prime}\bar{q^{\prime}}g$ \\
~ & with $q=u,\bar{u},d,\bar{d},s,\bar{s}$ \\
~ & and $q^{\prime}=u,d,s$ \\
~ & and $q,q^{\prime}$ not in the same flavor \\\hline
\chCode{}
u u > cc~(3S11) u u g
\end{lstlisting} & $qq\rightarrow c\bar{c}[\ss]+qqg$ \\
~ & with $q=u,\bar{u},d,\bar{d},s,\bar{s}$ \\\hline
\chCode{}
u d > cc~(3S11) u d g
\end{lstlisting} & $qq^{\prime}\rightarrow c\bar{c}[\ss]+qq^{\prime}g$ \\
~ & with $q,q^{\prime}=u,\bar{u},d,\bar{d},s,\bar{s}$\\
~ & and $q,q^{\prime}$ not in the same flavor
\\
\hline\hline
\end{tabular}
\end{center}
\caption{\label{tab:nnlosubproc}Subprocesses are calculated with each generation for $pp\rightarrow \psi+$3-jets in CSM.}
\end{table}

We have compared the \HELACOnia\ result with the \MadOnia\ result, and found they were in perfect agreement. Because \HELACOnia\ is based on the recursion relations, \HELACOnia\ is faster than \MadOnia\ in the computations. It is much easier for us to extend the \MadOnia\ result to a wider $p_T$ range. In Ref.~\cite{Aad:2014fpa}, ATLAS Collaboration already used our result to compare their measurment for $\psi(2S)$ up to $100$ GeV. Here, we present the $p_T$ distributions of $J/\psi$ (Fig.\ref{fig:NNLOstarLHCb}) in the LHCb acceptance at 13 TeV. We should remind the reader the following points. The color-singlet LDME is estimated in potential model~\cite{Eichten:1995ch}. The corresponding radial wave function at the orgin was derived in the QCD-motivated Buchmuller-Tye potential~\cite{Buchmuller:1980su}.\footnote{We will use the same color-singlet LDME in the following computations.} We used CTEQ6M~\cite{Pumplin:2002vw} as our PDF set and fixed $s_{ij}^{\rm min}=2m_c$. The error bands in Fig.\ref{fig:NNLOstarLHCb} represent the renormalization and factorization scale uncertainties $\frac{\sqrt{(2m_c)^2+p_T^2}}{2}\le \mu_R=\mu_F\le 2\sqrt{(2m_c)^2+p_T^2}$ and the uncertainty in charm quark mass $m_c=1.5\pm 0.1$ GeV. In Fig.\ref{fig:NNLOstarLHCb}, we do not include the NLO contribution, although it should be contained for a real NNLO$^\star$ prediction.

\begin{figure}[t!]
\begin{center}
\includegraphics[width=0.9\textwidth,draft=false]{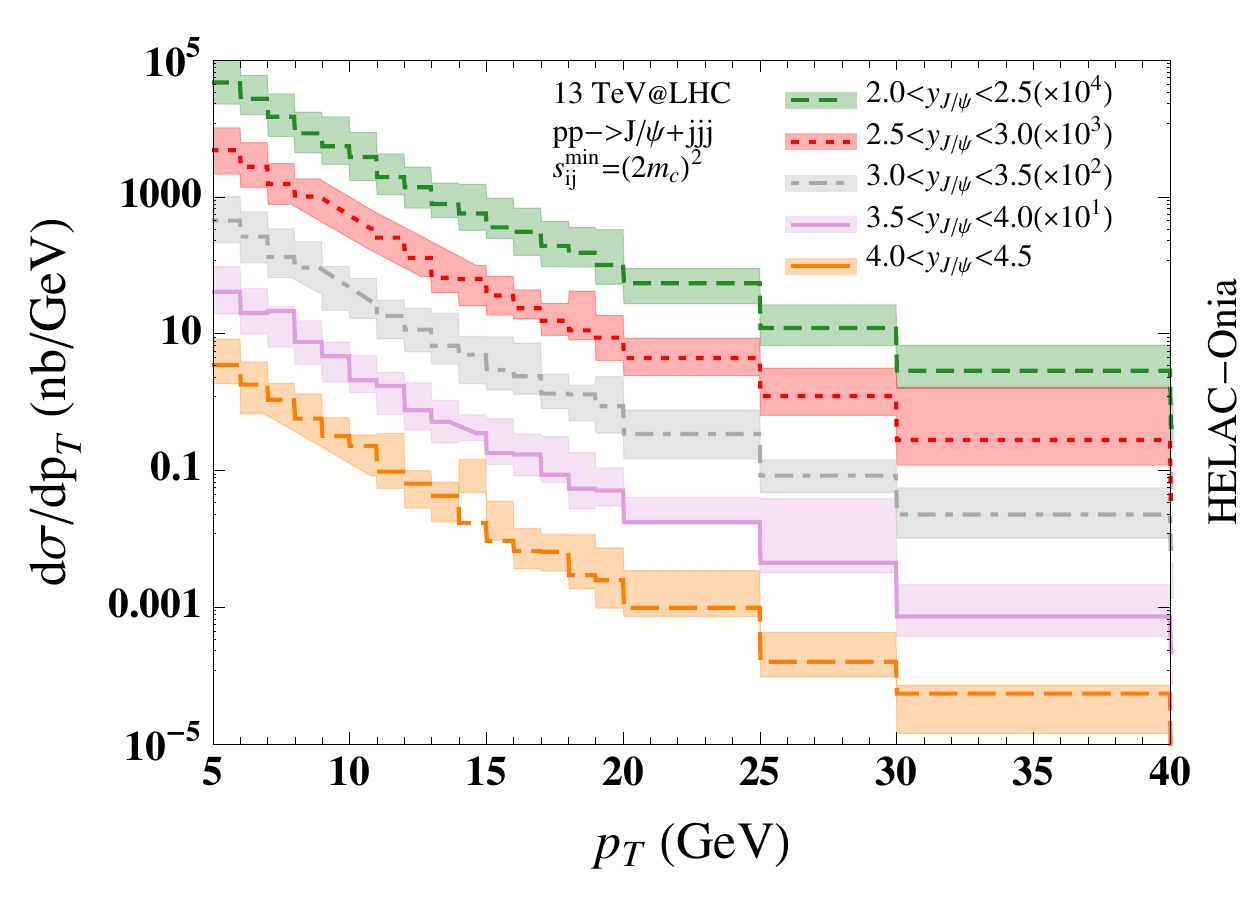}
\caption{The transverse momentum distributions of $J/\psi$ from $pp\to J/\psi+jjj$ in the LHCb acceptance at 13 TeV.
}
\label{fig:NNLOstarLHCb}
\end{center}\vspace*{-1cm}
\end{figure}

\subsection{$pp\to J/\psi+J/\psi+c\bar{c}$}

As noticed in Ref.~\cite{Lansberg:2014swa}, in double $J/\psi$ production, $pp\to J/\psi+J/\psi+c\bar{c}$ shares the leading-$p_T$ contribution from charm quark fragmentation diagrams though it is of $\mathcal{O}(\alpha_s^2)$ suppressed compared to $pp\to J/\psi+J/\psi$. Hence, it is necessary to quantify its maginitude by an explict calculation. For the first time, we performed such a complex calculation with the help of \HELACOnia\ 2.0 in Ref.~\cite{Lansberg:2014swa}, which involves more than 2000 Feynman diagrams. It is a first ( and till now the only ) $2\to 4$ process with at least two quarkonia to be calculated. We take this example to show the uniqueness and the rubostness of \HELACOnia\ to perform perturbative computations of more than one quarkonium production processes. Because the luminosity of the quark-antiquark initial states is usually much smaller than that of the gluon-gluon initial state at the high-energy colliders, we only include the gluon-gluon initial state here. One can use the following command line to generate the process
\cCode{}
HO> generate g g > cc~(3S11) cc~(3S11) c c~
\end{lstlisting}

For illustration, we are working in the CMS acceptance~\cite{Khachatryan:2014iia}:
\bqa
p_T^{J/\psi}> 6.5~{\rm GeV}~~~~~~~~~~~~~~~~~&{\rm if}&~~~~~~~~~~~~~~~|y^{J/\psi}|<1.2,\nonumber\\
p_T^{J/\psi}> 6.5 \rightarrow 4.5~{\rm GeV}~~~~~~~~&{\rm if}&~~~~~~~1.2<|y^{J/\psi}|<1.43,\nonumber\\
p_T^{J/\psi}> 4.5~{\rm GeV}~~~~~~~~~~~~~~~~~&{\rm if}&~~~~~~1.43<|y^{J/\psi}|<2.2,\label{CMSacc}
\eqa
where in the rapidity interval $1.2<|y^{J/\psi}|<1.43$, the transverse momentum $p_T^{J/\psi}$ cutoff scales linearly with its absolute rapidity $|y^{J/\psi}|$. We used the exact setup in Ref.~\cite{Lansberg:2014swa}. Some selected differential distributions in $pp\to J/\psi+J/\psi+c\bar{c}$ are shown in Fig.\ref{fig:psipsiccbarCMS}. The error bands are coming from the variations of the renormalization and factorization scales and the uncertainty of charm quark mass. Following the way in Ref.~\cite{Lansberg:2014swa}, we have taken into account the feeddown contribution from $\psi(2S)$ decay. The feeddown contribution  enhances the (differential) cross section by a factor of $1.89$. The absolute azimuthal difference between the $J/\psi$ pair $\frac{d\sigma}{d\Delta \phi}$ is shown in Fig.\ref{fig:dsigCMSa}. It is an observable to distuiguish the double-parton scattering (DPS) and the traditional single-parton scattering (SPS) since in the former production mechanism the two $J/\psi$ are uncorrelated. However, such an observable might be sensitive to the primordial $k_T$ smearing from the beam~\cite{Lansberg:2013qka}. Besides the absolute azimuthal difference, the absolute rapidity difference (Fig.\ref{fig:dsigCMSb}) and the invariant mass distribution (Fig.\ref{fig:dsigCMSc}) are also good kinematical variables to discriminate DPS and SPS. Finally, various transverse momentum spectra are displayed. In Fig.\ref{fig:dsigCMSd}, we presented distribution of the vectorial transverse momentum sum $P_T^{J/\psi J/\psi}=|\vec{p}_{T1}^{J/\psi}+\vec{p}_{T2}^{J/\psi}|$, while Fig.\ref{fig:dsigCMSe} (Fig.\ref{fig:dsigCMSf}) is the yileds of the leading $p_T={\rm max}(p_{T1}^{J/\psi},p_{T2}^{J/\psi})$ (subleading $p_T={\rm min}(p_{T1}^{J/\psi},p_{T2}^{J/\psi})$) of the two $J/\psi$. 

\begin{figure}[t!]
\begin{center}
\subfloat[Absolute azimuthal difference between $J/\psi$ pair]{\includegraphics[width=0.49\textwidth,draft=false]{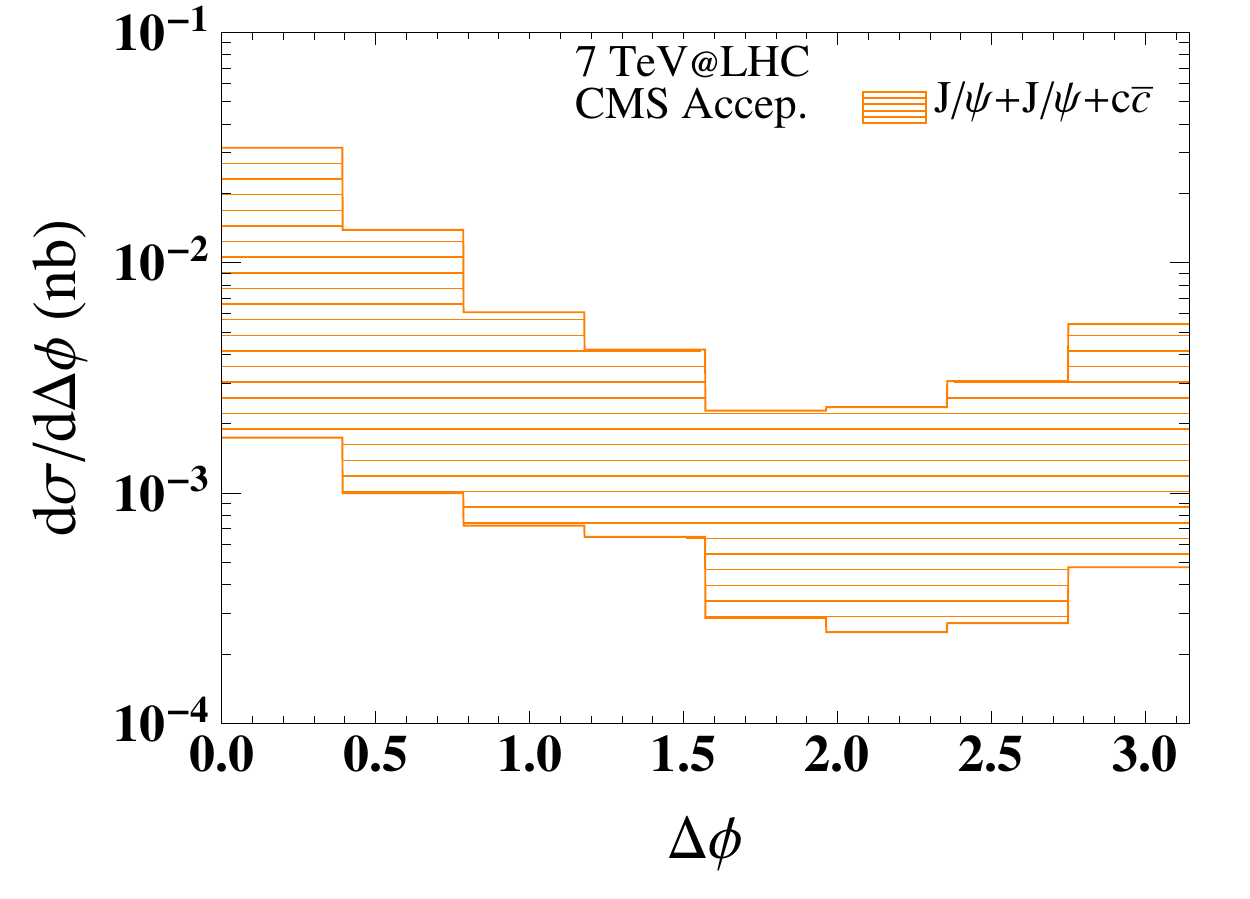}\label{fig:dsigCMSa}}
\subfloat[Absolute rapidity difference between $J/\psi$ pair]{\includegraphics[width=0.49\textwidth,draft=false]{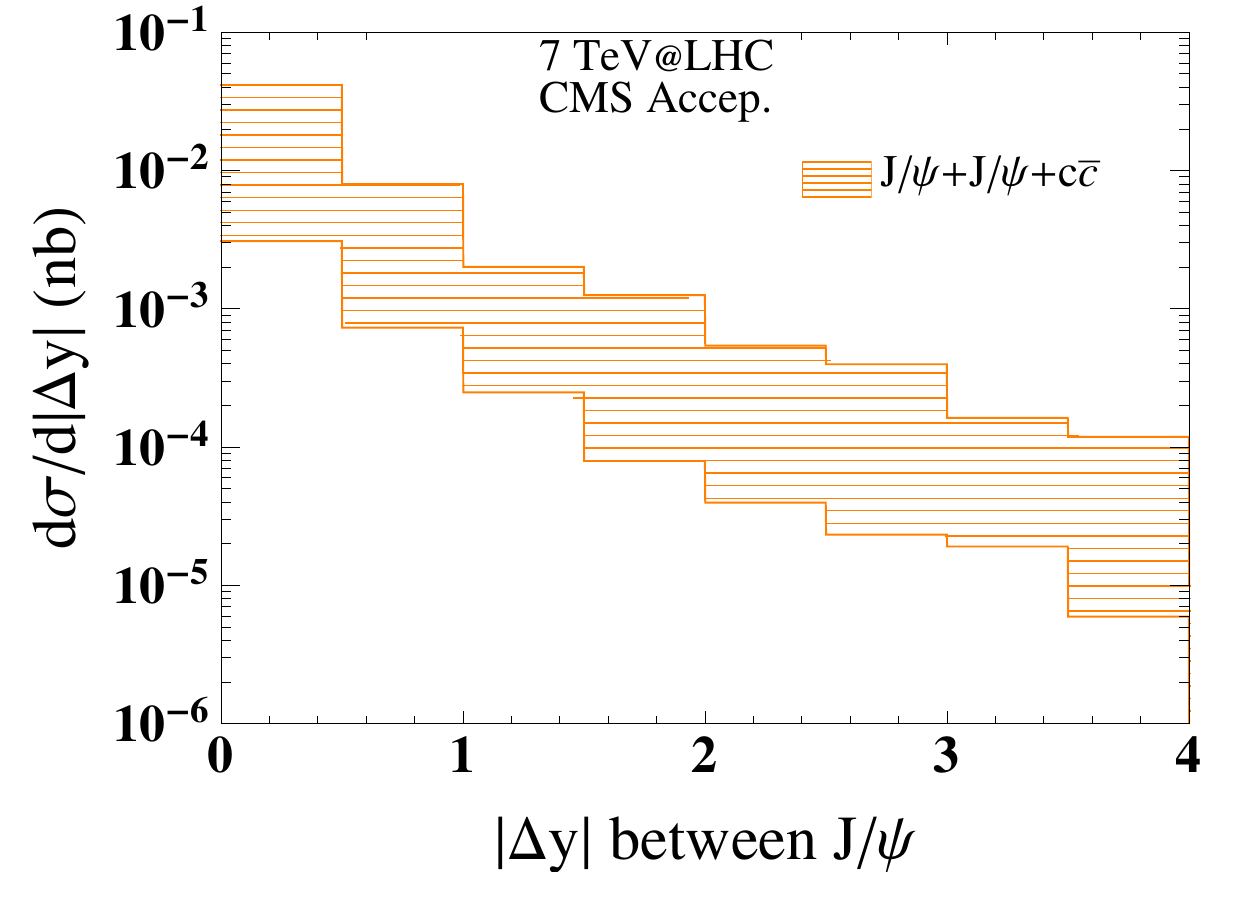}\label{fig:dsigCMSb}}\\
\subfloat[Invariant mass distribution]{\includegraphics[width=0.49\textwidth,draft=false]{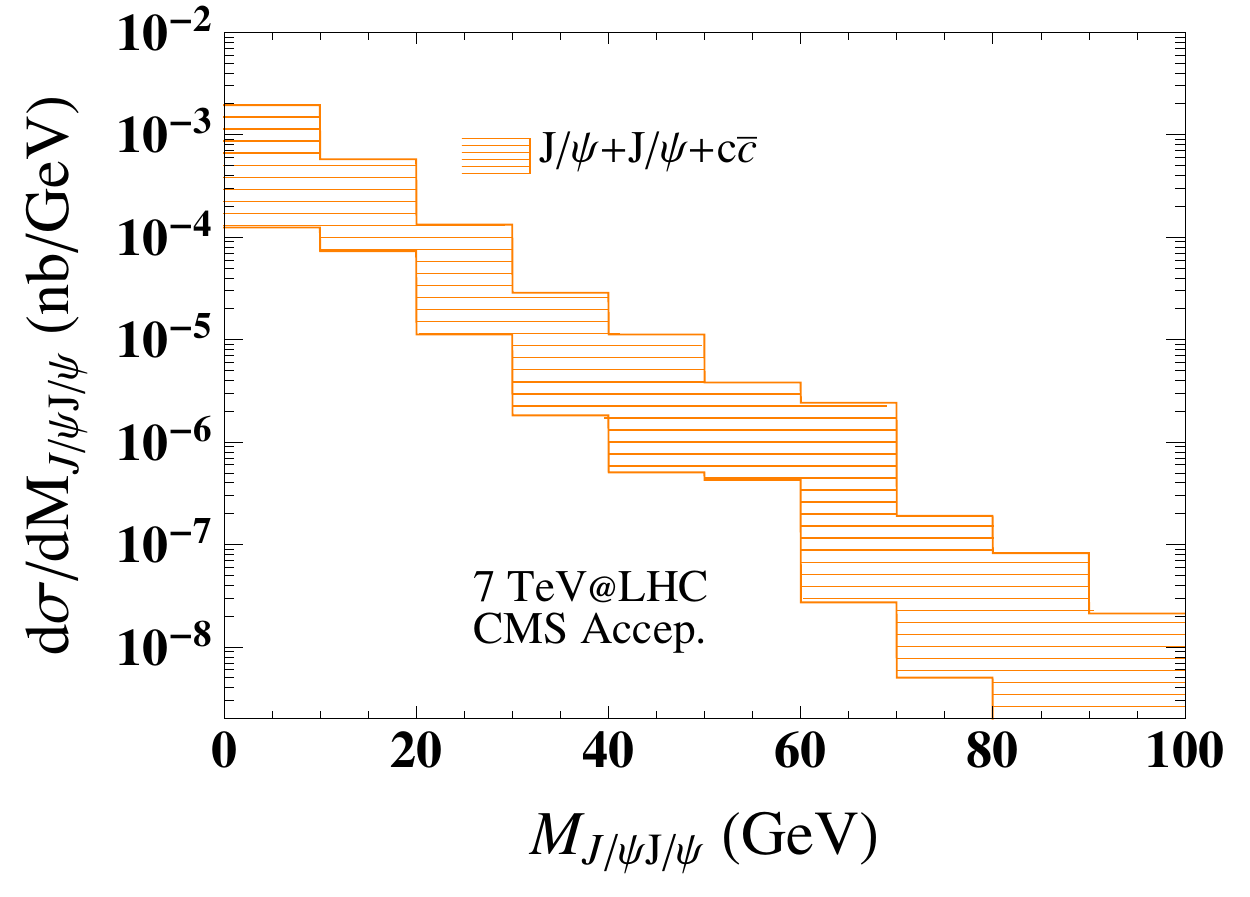}\label{fig:dsigCMSc}}
\subfloat[$P_T^{J/\psi J/\psi}$ distribution]{\includegraphics[width=0.49\textwidth,draft=false]{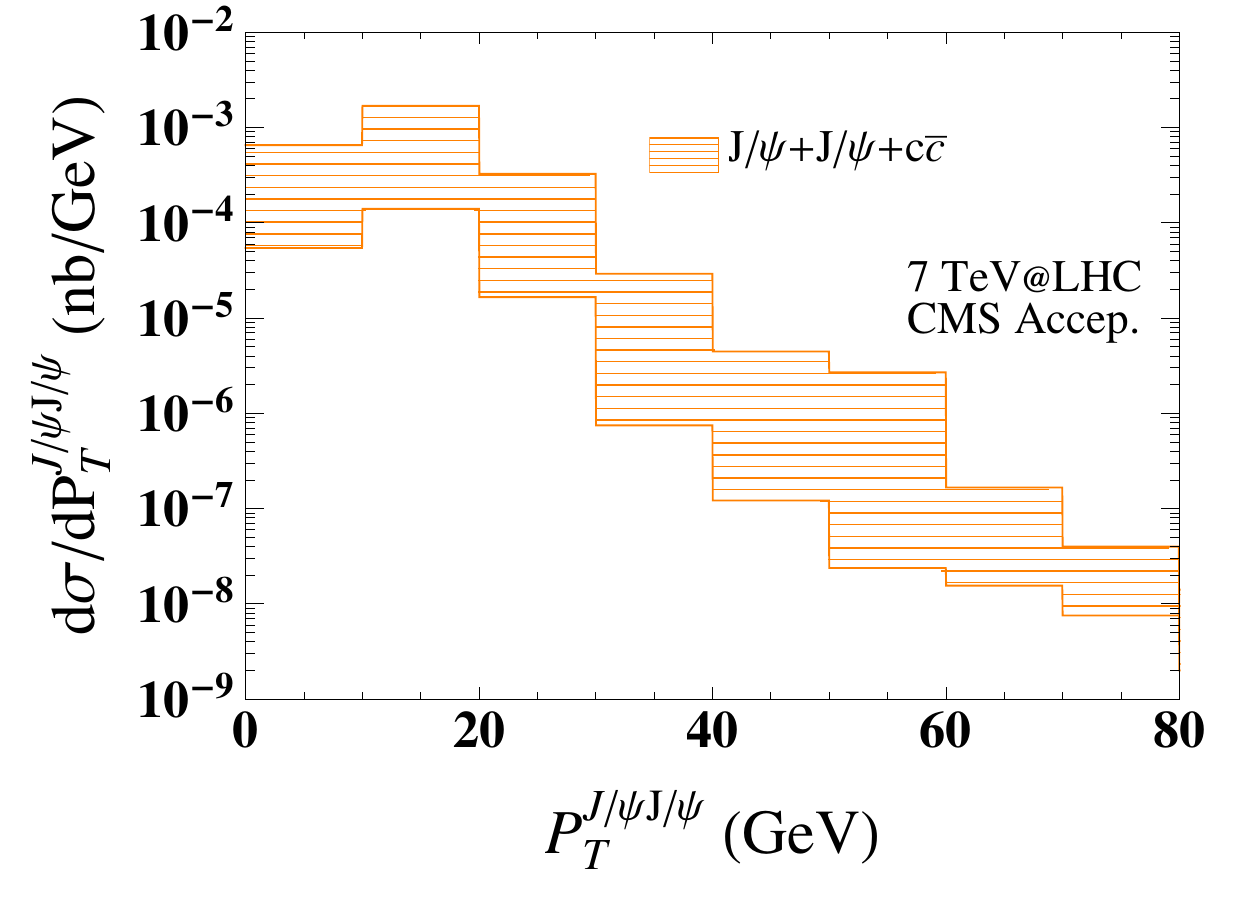}\label{fig:dsigCMSd}}\\
\subfloat[Leading $p_T$ between $J/\psi$ pair]{\includegraphics[width=0.49\textwidth,draft=false]{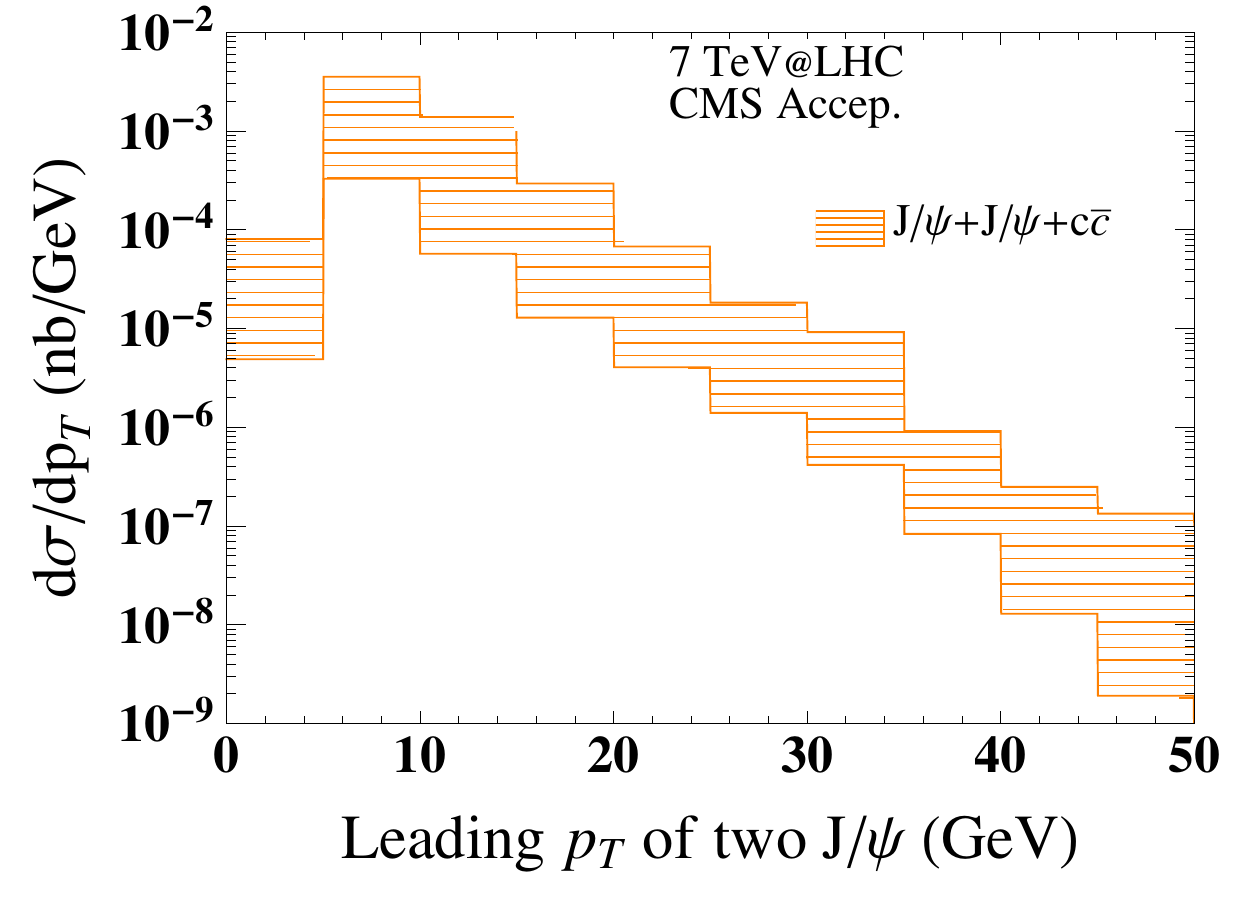}\label{fig:dsigCMSe}}
\subfloat[Sub-leading $p_T$ between $J/\psi$ pair]{\includegraphics[width=0.49\textwidth,draft=false]{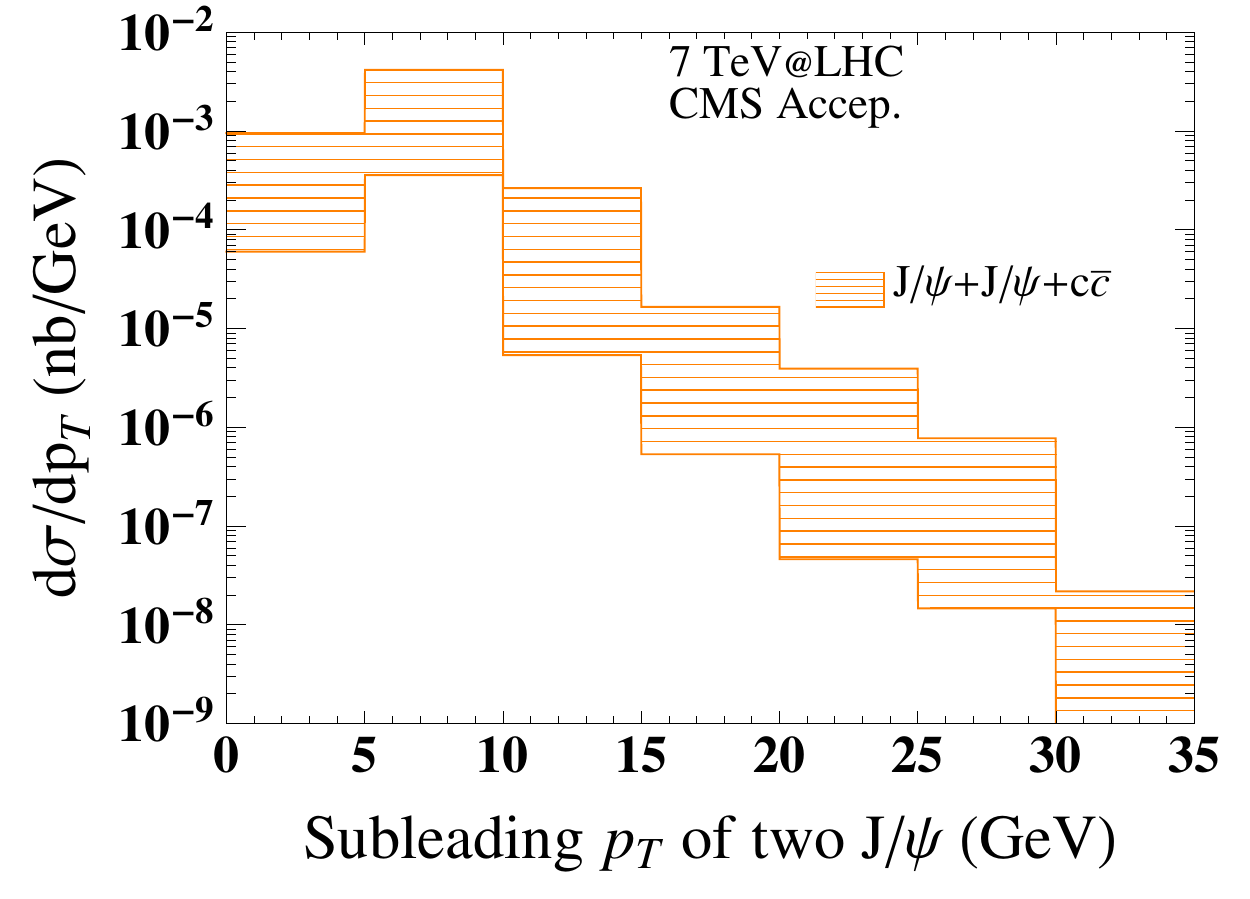}\label{fig:dsigCMSf}}
\caption{The differential distributions for $pp\to J/\psi+J/\psi+c\bar{c}$ in the CMS fidicuial region~\cite{Khachatryan:2014iia}:(a) absolute azimutal difference; (b) absolute rapidity difference ; (c) invariant mass distribution; (d) the vectorial transverse momentum sum; (e) leading $p_T$;
(f) sub-leading $p_T$.
}
\label{fig:psipsiccbarCMS}
\end{center}\vspace*{-1cm}
\end{figure}

\subsection{$J/\psi$ hadroproduction with parton shower effect}

The inclusive $J/\psi$ hadroproduction is a first process challenging our understanding of the heavy-quarkonium-production mechanism. For a long time, it was known that the CSM can describe the total cross section of $J/\psi$ or $\Upsilon$ (e.g. see Ref.~\cite{Brambilla:2004wf}) but not in the transverse momentum $p_T$ distributions. In the recent years, most of the studies were focusing on the interpretation of the yields~\cite{Campbell:2007ws,Artoisenet:2008fc,Butenschoen:2011yh,Butenschoen:2010rq,Ma:2010yw,Ma:2010jj,Han:2014jya,Zhang:2014ybe,Butenschoen:2014dra,Ma:2010vd} and the polarization~\cite{Gong:2008zz,Butenschoen:2012px,Chao:2012iv,Shao:2012fs,Gong:2012ug,Shao:2014yta,Shao:2014fca,Han:2014kxa,Faccioli:2014cqa,Bodwin:2014gia} of single-quarkonium hadroproduction at the large $p_T$ regime. Some efforts have also been paid in the small $p_T$ regime~\cite{Sun:2012vc,Ma:2014mri}. However, none of the consistent matching between large $p_T$ and small $p_T$ results exists. In Ref.~\cite{Sun:2012vc}, analytical small $p_T$ resummation is performed for the color-octet states only in NRQCD, which lacks the dominant color-singlet contribution and the matching to the fixed-order results. Alternatively, one can perform a resummation with the parton shower (PS) approach, which is formally to be restricted to the leading-log accuracy (although the partial subleading-log contributions can also be taken into account). It generates the complete events with correct  kinematics and can be applied directly on the experimental analysis by including the detector effect.

In this subsection, we will give a simple example to show the importance of parton shower effect for $J/\psi$ hadroproduction in the small $p_T$ region. Let us consider the color-singlet contribution only at leading order (LO) in $\alpha_s$ and in $v$ (the relative velocity between the charm quark pair). Without primordial $k_T$ smearing effect from the beam and the multiple interactions, the LO curve in $p_T$ distribution is indeed siginificantly smearing by PS as seen in Fig.\ref{fig:jpsips}, where we have used \Pythia8.186. In the left pannel of Fig.\ref{fig:jpsips}, one can observe that such a smearing effect is mainly from the initial state radiation (ISR) while the final state radiation (FSR) only distorts the distribution slightly. On the other hand,LO result is good enough to describe the rapidity distribution. We should remind the reader that while the LO+PS\footnote{One should turn on primordial $k_T$ in \Pythia\ as well.} color-singlet contribution is expect to describe the small $p_T$ data, the intermediate and the large $p_T$ data will receive substaintial higher-order (or real emissions)~\cite{Campbell:2007ws,Artoisenet:2008fc} and color-octet contributions. A consistent treatment of $J/\psi$ production in NRQCD is possible with the LO {\it merging} of matrix elements and PS in different jet multiplicities~\cite{Mrenna:2003if,Alwall:2007fs}, where a pioneer work has been done in $\eta_b$ production~\cite{Maltoni:2004hv}. Such a detailed analysis is of course interesting but beyond the scope of this paper. We will leave it for a future work.

\begin{figure}
\begin{center}
\vspace{-2cm}\hspace{-1cm}\includegraphics[width=0.5\textwidth]{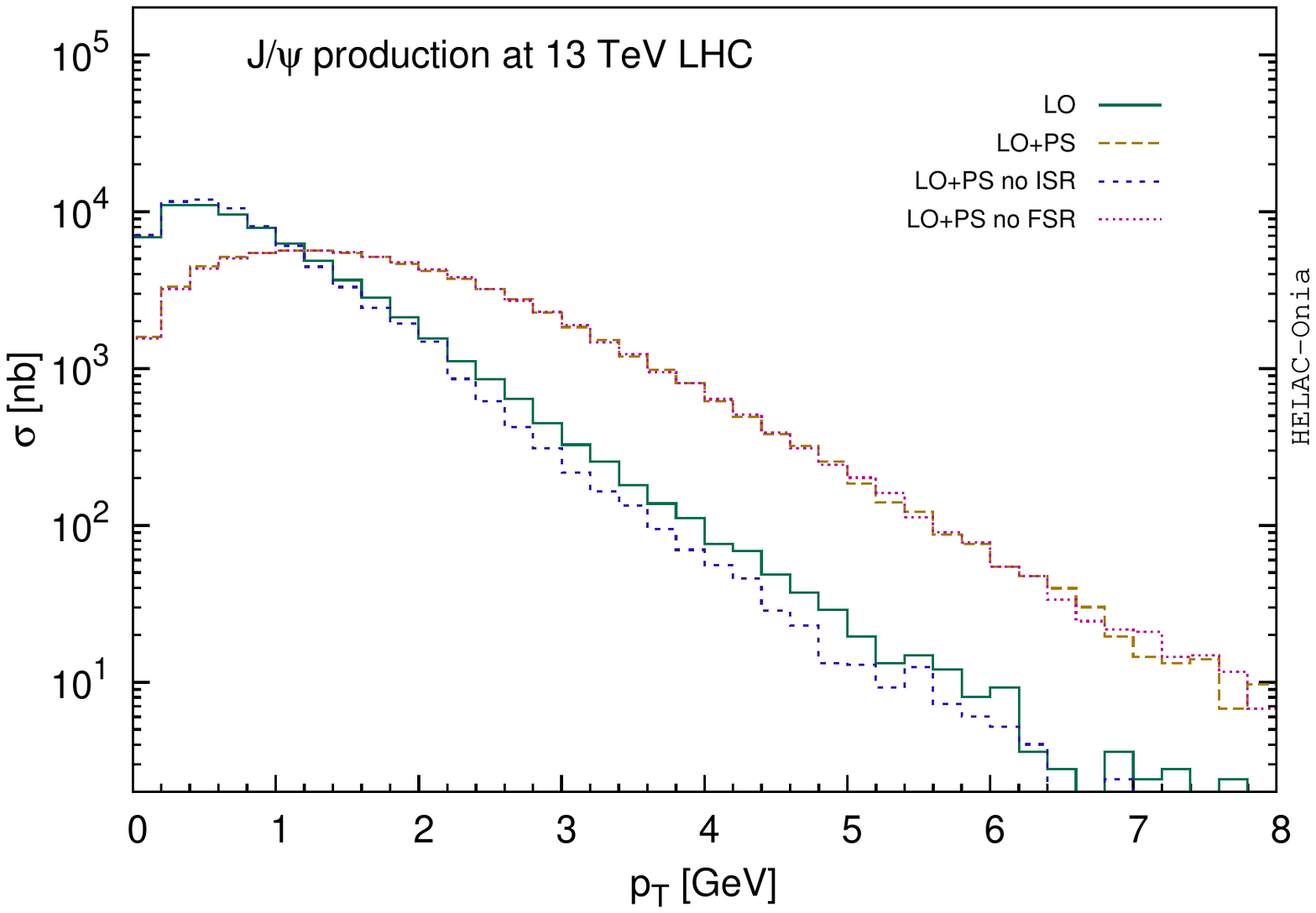}
\hspace{-0.75cm}\includegraphics[width=0.5\textwidth]{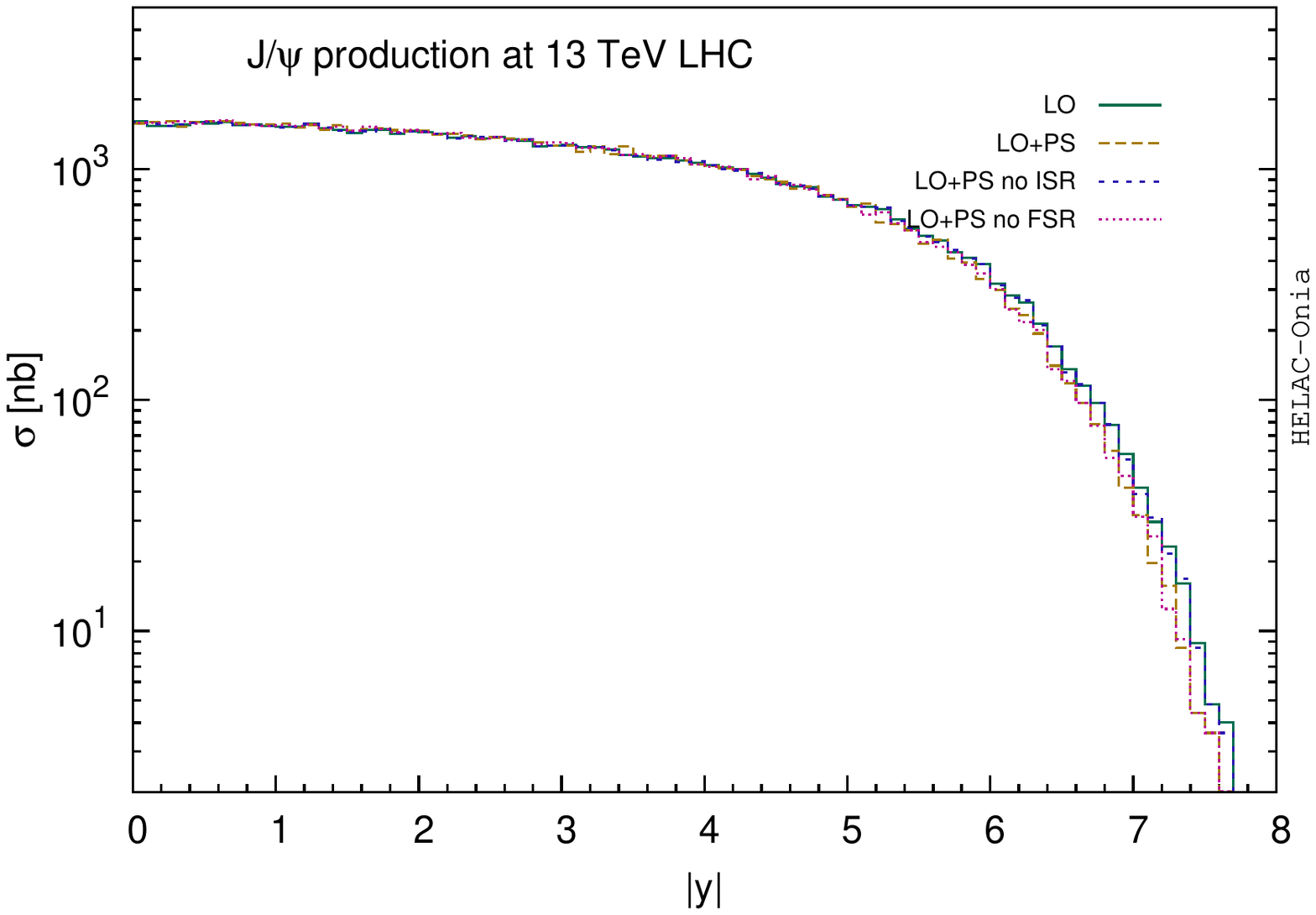}
\caption{\label{fig:jpsips}Illustrative plots for $J/\psi$ production at 13 TeV LHC with parton shower from \Pythia8.186. We presented the fixed-order LO calculation (solid curve), LO+PS (long-dashed curve), LO+PS but turning off ISR (short-dashed curve) and LO+PS but turning off FSR (dotted curve).}
\end{center}
\end{figure}

\subsection{Validation of decay angular distributions}

In subsection \ref{subsec:angdis}, we have discussed the implementation of the angular distributions in the heavy-quarkonium decay in \HELACOnia. Here, we will give three examples to validate our implementations: $J/\psi\rightarrow \ell^+\ell^-$, $\chi_{c1} \rightarrow J/\psi+\gamma$ and $\chi_{c2} \rightarrow J/\psi+\gamma$.

In our first example $J/\psi \rightarrow \ell^+\ell^-$, we know the angular distribution of one lepton is in the form of
\bqa
\frac{d\sigma}{d\cos{\theta}} \sim 1+\lambda_{\theta}\cos^2{\theta},\label{eq:psi2mm}
\eqa
where $\theta$ is the polar angular respect to the spin quantization axis in the rest frame of $J/\psi$ and $\lambda_{\theta}$ can be expressed in the longitudinal polarized cross section $\sigma_{\rm L}$ and the transverse polarized cross section $\sigma_{\rm T}$ of $J/\psi$
\bqa
\lambda_{\theta}=\frac{\sigma_{\rm T}-2\sigma_{\rm L}}{\sigma_{\rm T}+2\sigma_{\rm L}}.
\eqa
We have compared the numerical result from \HELACOnia\ and the analytical result Eq.(\ref{eq:psi2mm}) in Fig.\ref{fig:angulardistribution1}. The total cross section has been normalized to unity. We selected $\lambda_{\theta}=1,-1,0,0.3,0.5$ for illustration. Perfect agreement is found.

For $\chi_{c1} \to J/\psi+\gamma$, we have the same polar angular distribution Eq.(\ref{eq:psi2mm}) with respect to the decay product $J/\psi$ or $\gamma$~\cite{Shao:2012fs}, while for $\chi_{c2} \to J/\psi+\gamma$ the general formula is
\bqa
\frac{d\sigma}{d\cos{\theta}} \sim 1+\lambda_{\theta}\cos^2{\theta}+\lambda_{2\theta}\cos^4{\theta}.\label{eq:chic22psia}
\eqa
In the later case, $\lambda_{2\theta}$ is suppressed by the higher-order multipole amplitudes~\cite{Shao:2012fs}. Explicitly, we have for $\chi_{c1}$
\bqa
\lambda_{\theta}=(1-3\delta)\frac{\sigma^{\chi_{c1}}_{\rm tot}-3\sigma_{0,0}^{\chi_{c1}}}{(1+\delta)\sigma^{\chi_{c1}}_{\rm tot}+(1-3\delta)\sigma_{0,0}^{\chi_{c1}}},\label{eq:lambdachic1}
\eqa
and for $\chi_{c2}$
\bqa
\lambda_{\theta}&=&6\left[(1-3\delta_0-\delta_1)\sigma_{\rm tot}^{\chi_{c2}}-(1-7\delta_0+\delta_1)(\sigma^{\chi_{c2}}_{1,1}+\sigma^{\chi_{c2}}_{-1,-1})-(3-\delta_0-7\delta_1)\sigma^{\chi_{c2}}_{0,0}\right]/R,\nonumber\\
\lambda_{2\theta}&=&(1+5\delta_0-5\delta_1)\left[\sigma_{\rm tot}^{\chi_{c2}}-5(\sigma^{\chi_{c2}}_{1,1}+\sigma^{\chi_{c2}}_{-1,-1})+5\sigma^{\chi_{c2}}_{0,0}\right]/R,\nonumber\\
R&\equiv& (1+5\delta_0+3\delta_1)\sigma^{\chi_{c2}}_{\rm tot}+3(1-3\delta_0-\delta_1)(\sigma^{\chi_{c2}}_{1,1}+\sigma^{\chi_{c2}}_{-1,-1})+(5-7\delta_0-9\delta_1)\sigma^{\chi_{c2}}_{0,0}.\label{eq:lambdachic2}
\eqa
In the above equation, we denote $\sigma_{i,j}^{\chi_{c1,2}}$ is the $(i,j)$-component of the spin density matrix of $\chi_{c1,2}$ production. In the following, we take $\sigma_{-i,-j}^{\chi_{c1,2}}=\sigma_{i,j}^{\chi_{c1,2}}$, which is valid in a CP-conserved process. 
The spin-summed cross sections can be expressed as
\bqa
\sigma^{\chi_{c1}}_{\rm tot}&=&\sigma^{\chi_{c1}}_{1,1}+\sigma^{\chi_{c1}}_{0,0}+\sigma^{\chi_{c1}}_{-1,-1},\nonumber\\
\sigma^{\chi_{c2}}_{\rm tot}&=&\sigma^{\chi_{c2}}_{2,2}+\sigma^{\chi_{c2}}_{1,1}+\sigma^{\chi_{c2}}_{0,0}+\sigma^{\chi_{c2}}_{-1,-1}+\sigma^{\chi_{c2}}_{-2,-2}.
\eqa
Parameters $\delta$,$\delta_0$,$\delta_1$ enter into Eq.(\ref{eq:lambdachic1}) and Eq.(\ref{eq:lambdachic2}). They can be determined by the normalized\footnote{We have $(a_1^{J=1})^2+(a_2^{J=1})^2=1$ and $(a_1^{J=2})^2+(a_2^{J=2})^2+(a_3^{J=2})^2=1$.} multipole amplitudes
\bqa
\delta&=&\frac{(1+2a_1^{J=1}a_2^{J=1})}{2},\nonumber\\
\delta_0&=&\frac{1+2a_1^{J=2}(\sqrt{5}a_2^{J=2}+2a_3^{J=2})+4a_2^{J=2}(a_2^{J=2}+\sqrt{5}a_3^{J=2})+3\left(a_3^{J=2}\right)^2}{10},\nonumber\\
\delta_1&=&\frac{9+6a_1^{J=2}(\sqrt{5}a_2^{J=2}-4a_3^{J=2})-4a_2^{J=2}(a_2^{J=2}+2\sqrt{5}a_3^{J=2})+7\left(a_3^{J=2}\right)^2}{30},
\eqa
where $a_{1}^{J=j}$,$a_{2}^{J=j}$ and $a_{3}^{J=j}$ are the electric dipole (E1), magnetic quadrupole (M2) and electric octupole (E3) amplitudes for $\chi_{cj}$. We take the measured values by CLEO collaboration in Ref.~\cite{Artuso:2009aa}. The numercial values are shown in Tab.\ref{tab:multipole}. However, the implementation of the cascade decay $\chi_c\to J/\psi\gamma\to \ell^+\ell^-\gamma$ in \HELACOnia\ requires its full knowledge of the helicity decay amplitudes in terms of multipole amplitudes. Its complete derivation was performed in Ref.~\cite{Shao:2014yfa} at the amplitude level for the first time. Such amplitude-level experssions will be served as the helicity amplitude $\mathcal{A}({\bf x})$ defined in subsection \ref{subsec:angdis}. The validation of the implementations for $\chi_{c1,2}\to J/\psi+\gamma$ in \HELACOnia\ 2.0 can be found in Fig.\ref{fig:angulardistribution2} and Fig.\ref{fig:angulardistribution3}. The histograms of the decay product $J\psi$'s angular distributions perfectly agree with the analytical experssions.

\begin{table}
\begin{center}
\begin{tabular}{{c|}*4 c} 
\hline\hline
~ & $a^{J=j}_{1}$ & $a^{J=j}_{2}$ & $a^{J=j}_{3}$ \\\hline
$j=1$ & $0.998$ & $-0.0626$ & -\\
$j=2$ & $0.996$ & $-0.093$ & $0$
\\
\hline\hline
\end{tabular}
\end{center}
\caption{\label{tab:multipole}The normalized multipole amplitudes of $\chi_{cj}\rightarrow J/\psi+\gamma$ from the CELO measurment~\cite{Artuso:2009aa}.}
\end{table}

\begin{figure}
\begin{center}
\includegraphics[width=0.99\textwidth]{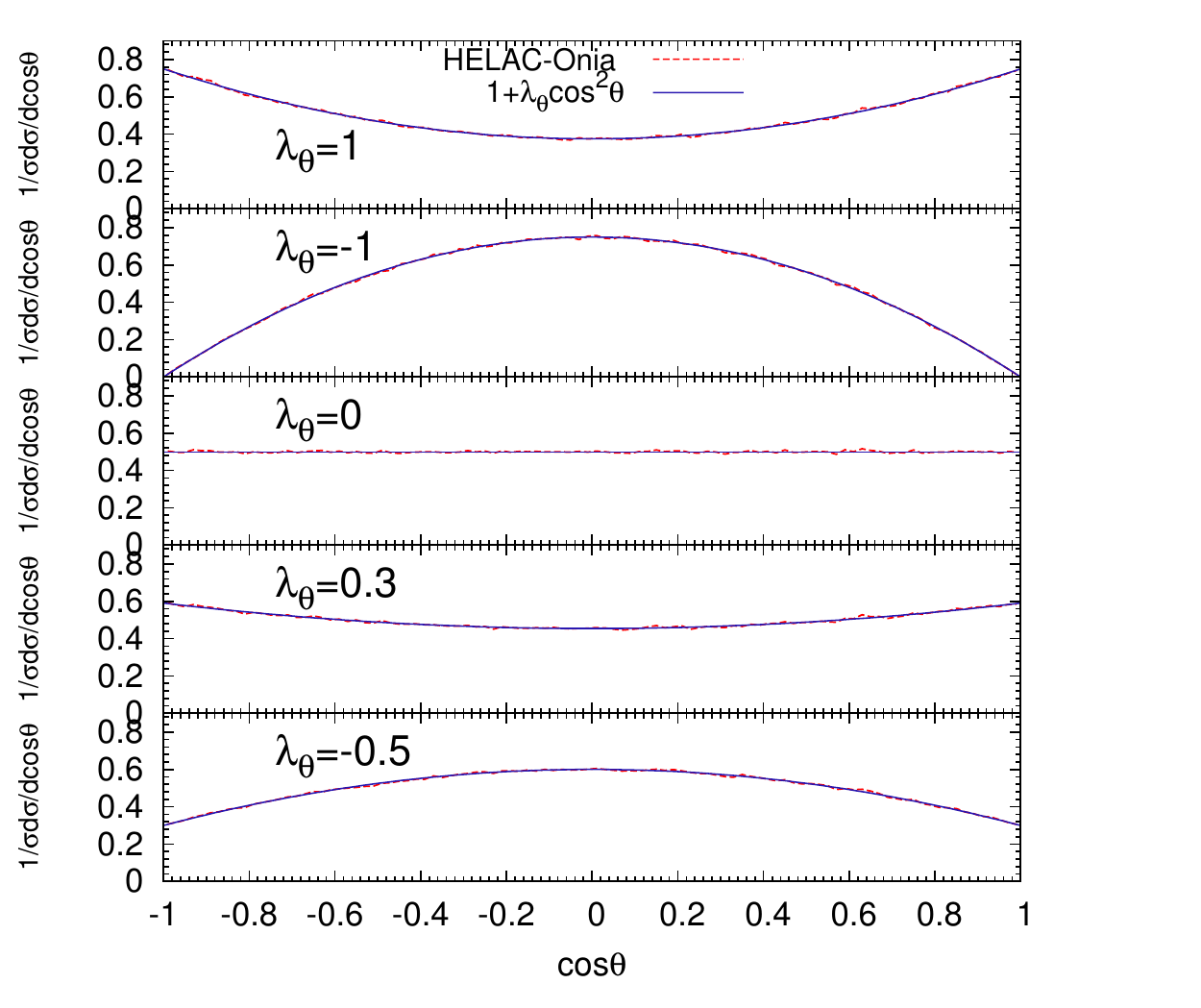}
\caption{\label{fig:angulardistribution1}Validation of lepton angular distributions in $J/\psi \rightarrow \ell^+\ell^-$.}
\end{center}
\end{figure}

\begin{figure}
\begin{center}
\includegraphics[width=0.99\textwidth]{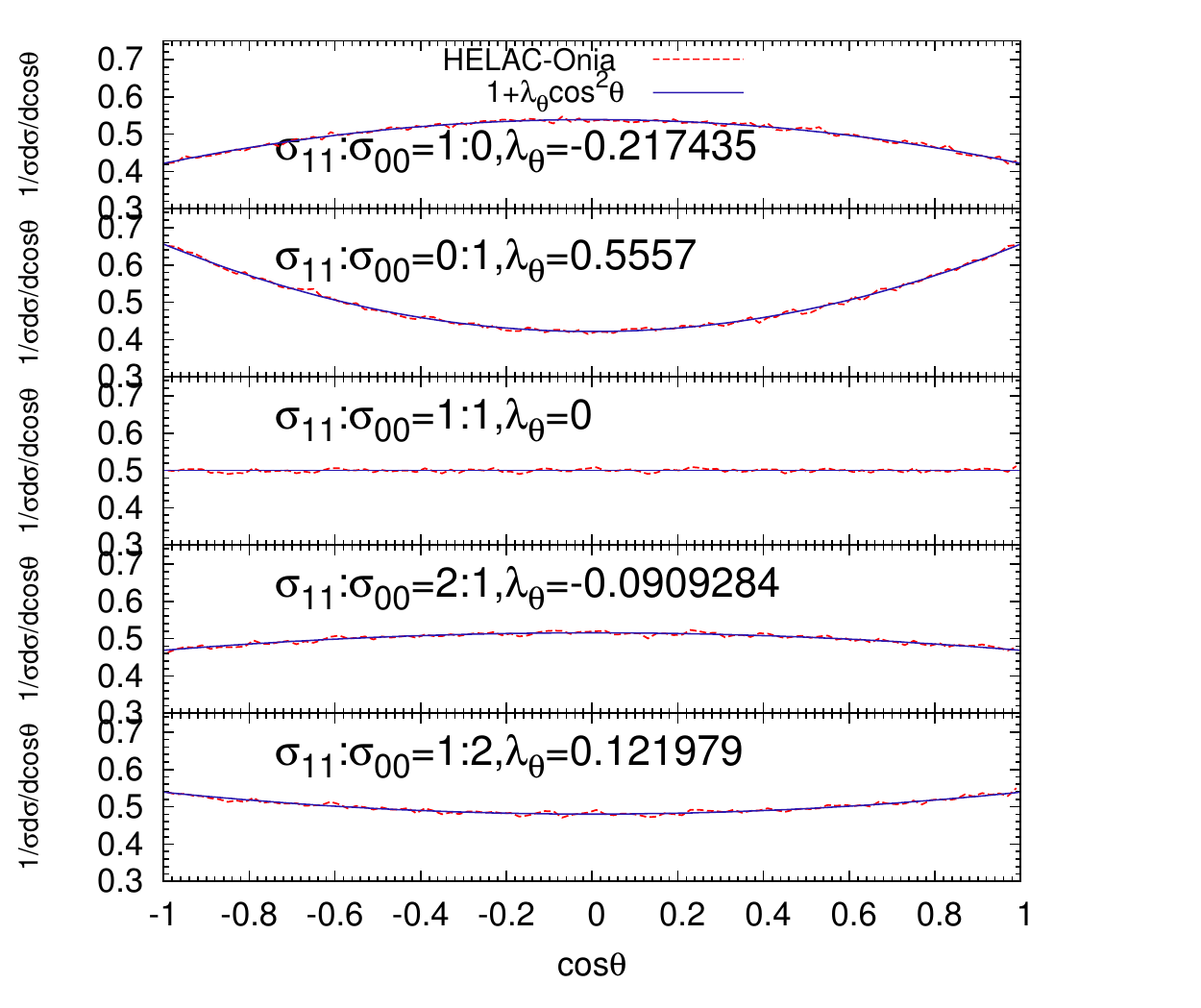}
\caption{\label{fig:angulardistribution2}Validation of $J/\psi$ angular distributions in $\chi_{c1} \rightarrow J/\psi+\gamma$.}
\end{center}
\end{figure}

\begin{figure}
\begin{center}
\includegraphics[width=0.99\textwidth]{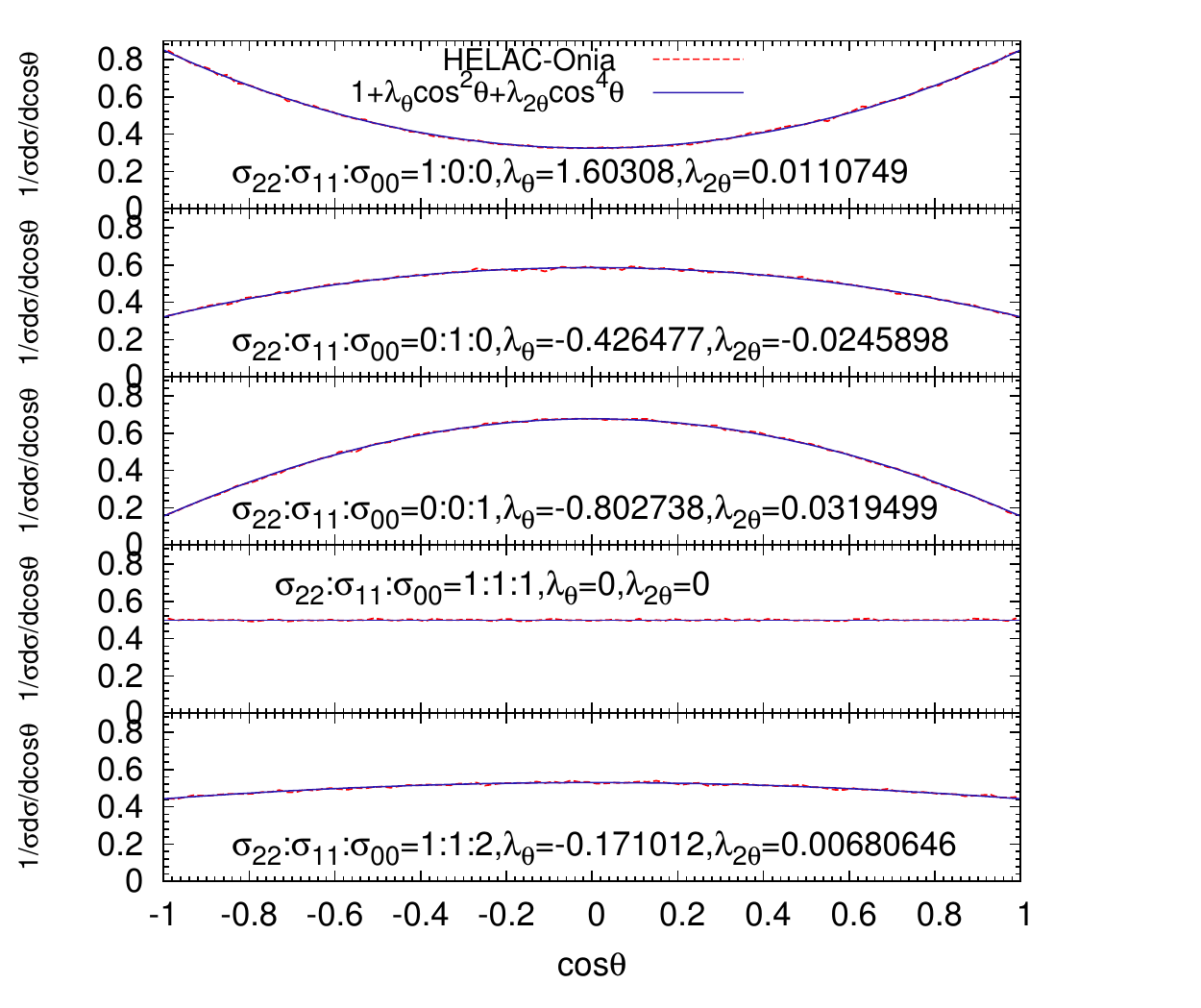}
\caption{\label{fig:angulardistribution3}Validation of $J/\psi$ angular distributions in $\chi_{c2} \rightarrow J/\psi+\gamma$.}
\end{center}
\end{figure}

\section{Conclusions \label{sec:con}}

We have presented a version 2.0 of \HELACOnia\ with several important updates for the practical theoretical studies and the Monte Carlo simulations for heavy-quarkonium-production processes. The main improvements are
\begin{itemize}
\item a completely new interface for talking between the user and the program written in \Python\ scripts. It is much user-friendly and  suitable to submit calculation jobs with multi-threading usage or on a cluster;
\item automated interfacing \HELACOnia\ to the parton shower Monte Carlo event generators. Two parton shower programs are sucessfully linked; One is QEDPS for the initial photon showering from the processes in electron-positron collisions, while the other one is the widely used one \Pythia\ 8;
\item a decay module for perfoming the spin-entangled (cascade-) decay of heavy quarkonium. Some dedicated decay processes are implemented such as $J/\psi \to \ell^+\ell^-$, $\chi_c\to J/\psi+\gamma$ and the decays of top quark, W-boson, Z boson;
\item a reweighting method for estimating the uncertainties from the renormalization/factorization scale and PDF in an automatic manner;
\item one-dimensional or two-dimensional histograms generation on the fly. Moreover, we also provide several useful analysis tools. 
\end{itemize}
All of the above improvements are quite useful in the study of the heavy-quarkonium production. It also provides a flexible framework for the future developments like heavy-quarkonium production in heavy ion collisions~\cite{Ferreiro:2013pua} or in the transverse momentum factorization framework~\cite{Dunnen:2014eta}.

\vfill


%





\appendix

\section{Program structure\label{sec:program}}

In this section, we will describe briefly the new program structure of \HELACOnia\ 2.0 for the future developments. The main files contained in the program are already described in the \textbf{README} file of the tarball. The files in the program are mainly included in several subdirectories, which are displayed in Fig.\ref{fig:hostructure}. There are mainly ten subdirectories under the main directory of \HELACOnia. Let us explain them in somewhat detail:
\begin{itemize}
\item \textbf{input}. All of the input files that required by the program are contained in this subdirectory. They are:
\begin{itemize}
\item \textbf{user.inp}: a file for user to specify his/her input parameters.
\item \textbf{default.inp}: a file that includes all of the default values for the input parameters.
\item \textbf{process.inp}: a file for user to tell the program the process information.
\item \textbf{ho\_configuration.txt}: a configuration file for \HELACOnia.
\item \textbf{seed.input}: a seed for random number generator.
\item \textbf{shower\_card\_user.inp} (\textbf{shower\_card\_default.inp}): a user (default) card to use parton shower programs.
\item \textbf{decay\_param\_user.inp} (\textbf{decay\_param\_default.inp}): a list of user-defined (default) parameters for using in the decay module.
\item \textbf{decay\_user.inp} (\textbf{decay\_default.inp}): a file to specify decay chains in this card.
\end{itemize}
\item \textbf{output}. All of the output files will be generated here. Initially, it is empty.
\item \textbf{src}. It contains all of the main source files of the program. They can be mainly divided into two parts. One  part is for the matrix elements generator and the other part is for the phase space integration and events generation.
\begin{enumerate}
\item matrix elements generation.
\begin{itemize}
\item \textbf{Helac\_Global.f90}: It is a file which contains all of the global variables.
\item \textbf{Helac\_Func\_1.f90}: In it, many helper functions and subroutines are defined.
\item \textbf{alfas\_functions.f90}: Running of $\alpha_S$ which is used in MCFM~\cite{Campbell:2010ff}.
\item \textbf{Projectors.f90}: It is a file in which the Clebsch-Gordan coefficients are defined.
\item \textbf{Constants.90}: Several subroutines are defined for reading input parameters.
\item \textbf{SM\_FeynRule\_Helac.f90}:It contains all of the Feynman rules of the Standard Model.
\item \textbf{Feynman\_Helac.f90}: A useful subroutine is written in this file for reconstructing all Feynman diagrams.
\item \textbf{Helac\_wavef.90}:It is a file to define all of external wave functions.
\item \textbf{Helac\_pan2.f90}: Definition of vertices to be used in Helac\_pan1.f90.
\item \textbf{Helac\_pan1.f90}:Off-shell currents generation by using recursion relation.
\item \textbf{Helac\_master.f90}: It is a main file of computing helicity amplitudes.
\end{itemize}
\item phase space integration and events generation. It is based on several adapted Monte Carlo integration programs.
\begin{enumerate}
\item PHEGAS:
\begin{itemize}
\item \textbf{Phegas.f90}:It is an extensive version of PHEGAS~\cite{Papadopoulos:2000tt} to deal with quarkonium kinematics. It was rewritten in \Fortran\ 90.
\item \textbf{Phegas\_Choice.f90}: Some helper functions are defined here that will be used by \textbf{Phegas.f90}. 
\end{itemize}
\item VEGAS:
\begin{itemize}
\item \textbf{MC\_VEGAS.f90}: A \Fortran\ 90 version of VEGAS~\cite{Lepage:1977sw}.
\item \textbf{Func\_PSI.f90}: Some helper functions of phase space integration were written in this file.
\item \textbf{Colliders\_PSI1.f90}: Phase space integration with VEGAS for $2\to n (n\ge 1)$ at hadron colliders.
\item \textbf{Colliders\_PSI2.f90}: Phase space integration with VEGAS for $2\to n (n\ge 2)$ at electron-positron colliders.
\end{itemize}
\item MINT:
\begin{itemize}
\item \textbf{mint-integrator.f90}: It is a \Fortran\ 90 version of \Mint~\cite{Nason:2007vt}.
\end{itemize}
\item Internal \Fortran\ 90 PDF files:
\begin{itemize}
\item \textbf{CTEQ6PDF.f90}:CTEQ6 PDF~\cite{Pumplin:2002vw} file in \Fortran\ 90 version.
\item \textbf{Structf\_PDFs.f90}: A file for calling PDFs.
\end{itemize}
\item LHAPDF file:
\begin{itemize}
\item \textbf{Structf\_LHAPDF.f90}: A file for calling PDFs from LHAPDF~\cite{Whalley:2005nh}. User should specify ``lhapdf=/path/to/lhapdf-config" in \textbf{input}/\textbf{ho\_configuration.txt}.
\end{itemize}
\item Others:
\begin{itemize}
\item \textbf{Helac\_ranmar.f90}: A random number generation program \Ranmar\ in \Fortran\ 90.
\item \textbf{MC\_PARNI\_Weight.f90}: PARNI in \Fortran\ 90, but it is not used.
\item \textbf{MC\_RAMBO.f90}: \Rambo~\cite{Kleiss:1985gy} in \Fortran\ 90.
\item \textbf{MC\_Helac\_GRID.f90}: A grid file.
\item \textbf{Helac\_unwei.f90}: There are some subroutines for dealing with unweighted events in this file.
\item \textbf{ADAPT.f90}: It is for optimization by using adaption procedure.
\item \textbf{Phegas\_Durham.f90}: Durham in \Fortran\ 90. It can only be used to generate phase space points for massless external particles.
\item \textbf{MC\_Func.f90}: There are some helper functions and subroutines for Monte Carlo integrations.
\item \textbf{Kinetic\_Func.f90}: Some kinematical variables are defined in this file.
\item \textbf{Cuts\_Module.f90}: It is a file to provide the user to impose kinematical cutoff.
\item \textbf{KT\_Clustering.f90}: $k_T$ clustering and reweight factor for MLM merging~\cite{Mangano:2006rw,Alwall:2007fs}.
\item \textbf{setscale.f90}: It provides the user to specify his/her renormalization and factorization scales.
\item \textbf{setscale\_default.f90}: It is only a default \textbf{setscale.f90} file for backup.
\item \textbf{Helac\_histo.f90}: Histogram drawing file in HELAC.
\item \textbf{SinglePro.f90}: It is the main file for phase space integration and events generation.
\item \textbf{Summation\_Pro.f90}: A file for the summation mode, which is not used yet.
\item \textbf{unweight\_lhe.f90}: A file for writing out Les Houches events files.
\item \textbf{FO\_plot.f90}: A file for plotting fixed-order distributions. In this case, unweight events generation is not necessary.
\item \textbf{Main\_Test.f90}: The \Fortran\ 90 main program.
\end{itemize}
\end{enumerate}
\end{enumerate}
\item \textbf{pdf}. More extensive internal PDFs are located in this subdirectory.
\begin{itemize}
\item \textbf{pdf\_list.txt}: A summary of internal PDFs in \HELACOnia.
\item \textbf{make\_opts},\textbf{makefile\_pdf}: Files of makefile for the PDF related routines. A library \textbf{libpdf.a} will be generated in \textbf{lib} subdirectory.
\item \textbf{opendata.f}: A file in \Fortran\ 77 for opening PDF data.
\item \textbf{Partonx5.f}: Standalone \Fortran\ 77 Partonx function.
\item CTEQ files: They include \textbf{cteq3.f},\textbf{Ctq4Fn.f},\textbf{Ctq5Par.f},\textbf{Ctq5Pdf.f},\textbf{Ctq6Pdf.f}.
\item MRST files: They include \textbf{mrs98.f},\textbf{mrs98ht.f},\textbf{mrs98lo.f},\textbf{mrs99.f},\textbf{mrst2001.f},\textbf{jeppe02.f}.
\item \textbf{gsdpdf} file: They include GS09 dPDF files~\cite{Gaunt:2009re}.
\end{itemize}
\item \textbf{shower}. The subdirectory contains files for parton shower.
\begin{itemize}
\item \textbf{QEDPS}: It contains the files of QEDPS for ISR photon shower form initial $e^{\pm}$ beams.
\item \textbf{PYTHIA8}: \Pythia\ 8 subsubdirectory. It includes the main files for interfacing \HELACOnia\ to \Pythia\ 8 for showering.
\item \textbf{PYTHIA6}: \Pythia\ 6~\cite{Sjostrand:2006za} subsubdirectory. It will be used for the future development.
\item \textbf{HERWIG6}: \Herwig\ 6~\cite{Corcella:2000bw,Corcella:2002jc} subsubdirectory. It will be used for the future development.
\item \textbf{HERWIGPP}: \Herwig++~\cite{Bahr:2008pv,Bellm:2013lba} subsubdirectory. It will be used for the future development.
\item \textbf{interface}: Some interface files are included in this subsubdirectory. For example, \textbf{QEDPS\_interface.f90} is a file to interface \HELACOnia\ with QEDPS.
\end{itemize}
\item \textbf{analysis}. A subdirectory for perfroming analysis.
\begin{itemize}
\item \textbf{hbook}: \Hbook\ files (a simplified version written by M. Mangano) for plotting.
\item \textbf{user}: user defined plot files,like \textbf{plot\_user.f90}. Some examples are also given.
\item \textbf{PYTHIA8}: the analysis code for generating histograms or \Root\ trees by using \Pythia8 and \FastJet~\cite{Cacciari:2011ma} (or its core functionality \FJCore).
\item \textbf{heptoptagger}: the \HEPTopTagger~\cite{Plehn:2009rk,Plehn:2010st,Plehn:2011sj} source code for top quark tagging in the analysis stage.
\item \textbf{include}: some including files for example \textbf{HEPMC90.INC} for defining \HepMC~\cite{Dobbs:2001ck} common variables.
\item \textbf{various}: some useful tools at the analysis stage.
\item \textbf{TMVA}: some examples for using TMVA contained in \Root\ for multiply variable analysis.
\item \textbf{LesHouches}: some useful tools for dealing the Les Houches event files.
\item \textbf{HepMC}: a code to convert \HepMC~\cite{Dobbs:2001ck} file to histograms or \Root\ trees by using \FastJet\ or \FJCore.
\end{itemize}
\item \textbf{jets}. A subdirecotry containing jet related tools.
\begin{itemize}
\item \textbf{fastjet}: code for interfacing \FastJet\ to \HELACOnia.
\item \textbf{fjcore}: the source code of \FJCore\ as well as the interface code to \HELACOnia.
\item \textbf{merge}: the different multiplicity leading-order matrix elements and parton shower merging code.
\end{itemize}
\item \textbf{cernlib}. A subdirectory containing cernlib files.
\begin{itemize}
\item \textbf{minuit}: \Minuit~\cite{James:1975dr} source files.
\end{itemize}
\item \textbf{decay}. A subdirectory for decaying final state particles.
\begin{itemize}
\item \textbf{decay\_list.txt}: a list of available decay processes.
\item \textbf{Decay\_interface.f90}: the main decay file.
\item \textbf{DecayInfo.f90}: a file to read the decay information from \textbf{decay\_user.inp} in \textbf{input} subdirectory.
\item \textbf{HOVll.f90}: the angular distribution file for a vector decays into two leptons.
\item \textbf{HO\_chi2psia.f90}: the angular distribution file for $\chi$ particle decays into a $J^{\rm PC}=1^{--}$ quarkonium and a photon.
\item \textbf{HO\_t2bw.f90}: the angular distribution file for top quark decays into a bottom quark and a W boson.
\end{itemize}
\item \textbf{cluster}. A subdirectory containing \Python\ scripts.
\begin{itemize}
\item \textbf{create\_subdir.sh}: a bash shell script for creating subdirectories in the working directory. It is useful for running on the cluster or in the multi-core mode.
\item \textbf{bin}: A subsubdirectory that contains executable script file \textbf{ho\_cluster} after configurating and make.
\item \textbf{pythoncode}: A subsubdirectory that contains the python source codes.
\begin{itemize}
\item \textbf{cluster.py}: A file includes various cluster classes. 
\item \textbf{misc.py}: Helpful functions defined to perform routine administrative I/O tasks.
\item \textbf{coloring\_logging.py}: A file with logging color.
\item \textbf{extended\_cmd.py}: A file conntaining different extension of the cmd basic python library.
\item \textbf{files.py}: A file contains useful classes for dealing with file access.
\item \textbf{helaconia\_run\_interface.oy} and \textbf{helaconia\_interface.py}: A user-friendly command line interface to access \HELACOnia\ features.
\end{itemize}
\end{itemize}
\item \textbf{addon}. A subdirectory for some {\it ad hoc} codes.
\begin{itemize}
\item \textbf{addon\_process.dat}: A list of available addon processes.
\item \textbf{pp\_psipsi\_DPS}: An {\it ad hoc} code for DPS of $pp(\bar{p})\to \mathcal{Q}_1\mathcal{Q}_2+X$, where $\mathcal{Q}_i=J/\psi,\psi(2S),\Upsilon(1S),\Upsilon(2S),\Upsilon(3S)$.
\item \textbf{pp\_psiX\_CrystalBall}: An {\it ad hoc} code for $pp(\bar{p})\to \mathcal{Q}+X$ via crystal ball function, where $\mathcal{Q}=J/\psi,\psi(2S),\Upsilon(1S),\Upsilon(2S),\Upsilon(3S),\chi_{c0},\chi_{c1},\chi_{c2}$ and $\chi_{bJ}(nP)$ with $J=0,1,2$,$n=1,2,3$.
\item \textbf{fit\_pp\_psiX\_CrystalBall}: An {\it ad hoc} code for fitting crystal ball function to the experimental data of $pp(\bar{p})\to \mathcal{Q}+X$, where $\mathcal{Q}=J/\psi,\psi(2S)$.
\item \textbf{fit\_pp\_upsilonX\_CrystalBall}: An {\it ad hoc} code for fitting crystal ball function to the experimental data of $pp(\bar{p})\to \mathcal{Q}+X$, where $\mathcal{Q}=\Upsilon(1S),\Upsilon(2S),\Upsilon(3S)$.
\item \textbf{fit\_pp\_QQ\_CrystalBall}: An {\it ad hoc} code for fitting crystal ball function to the experimental data of $pp(\bar{p})\to Q+\bar{Q}$, where $Q$ is charm or bottom quark.
\item \textbf{pp\_QQ\_CrystalBall}: An {\it ad hoc} code for generating events of $pp(\bar{p})\to Q+\bar{Q}$ via crystal ball function.
\item \textbf{pp\_aajj\_DPS}: An event generator for producing $pp(\bar{p})\to \gamma\gamma+$dijet from DPS.
\end{itemize}
\end{itemize}
There are other subdirectories under the main directory. All generated libraries will be put in \textbf{lib} subdirectory. All module (object) files will be put in \textbf{mod} (\textbf{obj}) subdirectory. Executable files will be generated in the subdirectory \textbf{bin}.

\begin{figure}
\begin{center}
\hspace{0cm}\includegraphics[width=0.85\textwidth]{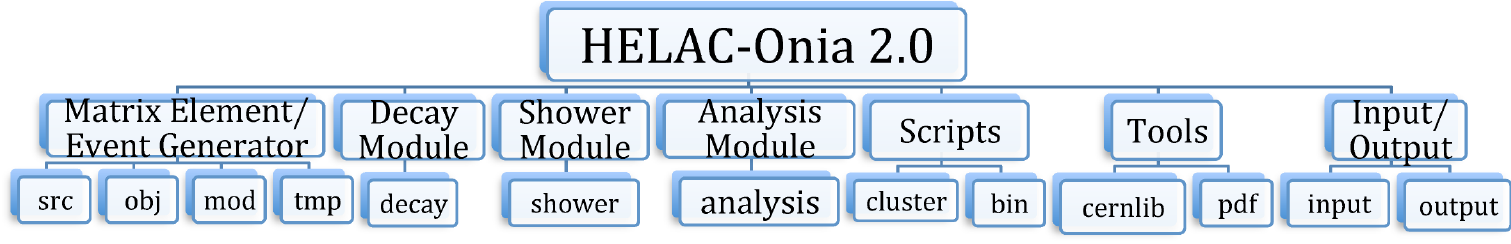}
\caption{\label{fig:hostructure}Program structure of \HELACOnia\ with version 2.0.}
\end{center}
\end{figure}

\section{Particle symbols in HELAC-Onia via Python script\label{app:symbol}}

In this appendix, we will introduce the new particle symbols for using \HELACOnia\ 2.0 with \Python\ scripts. We list them explicitly in Tabs.\ref{tab:smpart},\ref{tab:idcharmonia},\ref{tab:idbottomonia},\ref{tab:idBc}.

\begin{table}
\begin{center}
\begin{tabular}{c|c|c} 
\hline\hline
Particle & Particle ID & Particle Symbol \\
\hline
$\nu_e,e^-,u,d,\nu_\mu,\mu^-,c,s,\nu_{\tau},\tau^-,t,b$ & $1,\ldots,12$ & \ttfamily{ve},\ttfamily{e-},\ttfamily{u},\ttfamily{d},\ttfamily{vm},\ttfamily{m-},\ttfamily{c},\ttfamily{s},\ttfamily{vt},\ttfamily{tt-},\ttfamily{t},\ttfamily{b}
\\
 \multirow{2}{*}{$\bar{\nu_e},e^+,\bar{u},\bar{d},\bar{\nu}_\mu,\mu^+,
\bar{c},\bar{s},\bar{\nu_{\tau}},\tau^+,\bar{t},\bar{b}$} & \multirow{2}{*}{$-1,\ldots,-12$} & \ttfamily{ve$\sim$},\ttfamily{e+},\ttfamily{u$\sim$},\ttfamily{d$\sim$},\ttfamily{vm$\sim$},\ttfamily{m+},\\
~ & ~ & \ttfamily{c$\sim$},\ttfamily{s$\sim$},\ttfamily{vt$\sim$},\ttfamily{tt+},\ttfamily{t$\sim$},\ttfamily{b$\sim$}\\
$\gamma,Z,W^+,W^-,g$ & $31,\ldots,35$ & \ttfamily{a},\ttfamily{z},\ttfamily{w+},\ttfamily{w-},\ttfamily{g}\\
$H,\chi,\phi^+,\phi^-$ & $41,\ldots,44$ & \ttfamily{h},\ttfamily{g0},\ttfamily{g+},\ttfamily{g-}\\
\hline\hline
\end{tabular}
\end{center}
\caption{\label{tab:smpart}The identity numbers and symbols of the SM ``elementary" particles in \HELACOnia\ 2.0.}
\end{table}

\begin{table}
\begin{center}
\begin{tabular}{{c}*2 c} 
\hline\hline
Fock State & Particle ID & Particle Symbol\\\hline
$c\bar{c}[\oszs]$ & $441001$ & \ttfamily{cc$\sim$(1S01)}\\
$c\bar{c}[\oszo]$ & $441008$ & \ttfamily{cc$\sim$(1S08)}\\
$c\bar{c}[\tsos]$ & $443011$ & \ttfamily{cc$\sim$(3S11)}\\
$c\bar{c}[\tsoo]$ & $443018$ & \ttfamily{cc$\sim$(3S18)}\\
$c\bar{c}[\opos]$ & $441111$ & \ttfamily{cc$\sim$(3P11)}\\
$c\bar{c}[\opoo]$ & $441118$ & \ttfamily{cc$\sim$(3P18)}\\
$c\bar{c}[\tpjst]$ & $4431J1$ & \ttfamily{cc$\sim$(3PJ1)}\\
$c\bar{c}[\tpjot]$ & $4431J8$ & \ttfamily{cc$\sim$(3PJ8)}
\\
\hline\hline
\end{tabular}
\end{center}
\caption{\label{tab:idcharmonia}The identity numbers and symbols for the charmonia in various Fock states in \HELACOnia\ 2.0.}
\end{table}

\begin{table}
\begin{center}
\begin{tabular}{{c}*2 c} 
\hline\hline
Fock State & Particle ID & Particle Symbol\\\hline
$b\bar{b}[\oszs]$ & $551001$ & \ttfamily{bb$\sim$(1S01)}\\
$b\bar{b}[\oszo]$ & $551008$ & \ttfamily{bb$\sim$(1S08)}\\
$b\bar{b}[\tsos]$ & $553011$ & \ttfamily{bb$\sim$(3S11)}\\
$b\bar{b}[\tsoo]$ & $553018$ & \ttfamily{bb$\sim$(3S18)}\\
$b\bar{b}[\opos]$ & $551111$ & \ttfamily{bb$\sim$(3P11)}\\
$b\bar{b}[\opoo]$ & $551118$ & \ttfamily{bb$\sim$(3P18)}\\
$b\bar{b}[\tpjst]$ & $5531J1$ & \ttfamily{bb$\sim$(3PJ1)}\\
$b\bar{b}[\tpjot]$ & $5531J8$ & \ttfamily{bb$\sim$(3PJ8)}
\\
\hline\hline
\end{tabular}
\end{center}
\caption{\label{tab:idbottomonia}The identity numbers and symbols for the bottomonia in various Fock states in \HELACOnia\ 2.0.}
\end{table}

\begin{table}
\begin{center}
\begin{tabular}{{c}*2 c} 
\hline\hline
Fock State & Particle ID & Particle Symbol\\\hline
$c\bar{b}[\oszs]$ & $451001$ & \ttfamily{cb$\sim$(1S01)}\\
$c\bar{b}[\oszo]$ & $451008$ & \ttfamily{cb$\sim$(1S08)}\\
$c\bar{b}[\tsos]$ & $453011$ & \ttfamily{cb$\sim$(3S11)}\\
$c\bar{b}[\tsoo]$ & $453018$ & \ttfamily{cb$\sim$(3S18)}\\
$c\bar{b}[\opos]$ & $451111$ & \ttfamily{cb$\sim$(3P11)}\\
$c\bar{b}[\opoo]$ & $451118$ & \ttfamily{cb$\sim$(3P18)}\\
$c\bar{b}[\tpjst]$ & $4531J1$ & \ttfamily{cb$\sim$(3PJ1)}\\
$c\bar{b}[\tpjot]$ & $4531J8$ & \ttfamily{cb$\sim$(3PJ8)}\\\hline
$b\bar{c}[\oszs]$ & $-451001$ & \ttfamily{bc$\sim$(1S01)}\\
$b\bar{c}[\oszo]$ & $-451008$ & \ttfamily{bc$\sim$(1S08)}\\
$b\bar{c}[\tsos]$ & $-453011$ & \ttfamily{bc$\sim$(3S11)}\\
$b\bar{c}[\tsoo]$ & $-453018$ & \ttfamily{bc$\sim$(3S18)}\\
$b\bar{c}[\opos]$ & $-451111$ & \ttfamily{bc$\sim$(3P11)}\\
$b\bar{c}[\opoo]$ & $-451118$ & \ttfamily{bc$\sim$(3P18)}\\
$b\bar{c}[\tpjst]$ & $-4531J1$ & \ttfamily{bc$\sim$(3PJ1)}\\
$b\bar{c}[\tpjot]$ & $-4531J8$ & \ttfamily{bc$\sim$(3PJ8)}
\\
\hline\hline
\end{tabular}
\end{center}
\caption{\label{tab:idBc}The identity numbers and symbols for the mixed flavour quarkonium $B_c^{\pm}$ family in various Fock states in \HELACOnia\ 2.0.}
\end{table}

\section{New input parameters\label{app:param}}

Some of the parameters in \textbf{input}/\textbf{default.inp} and \textbf{input}/\textbf{user.inp} have already been introduced in Ref.~\cite{Shao:2012iz}. The new parameters we introduced in the new version are:
\begin{enumerate}
\item {\tt energy\_beam1} and {\tt energy\_beam2} are the energies in
unit of GeV of the first beam and second beam respectively.
\item {\tt fixtarget} is a flag to compute the cross section in a fixed-target collision envrioment (T) or not (F).
\item {\tt ranhel} is a parameter to determine whether the program uses the
Monte Carlo sampling over the helicity configurations. In \HELACOnia\ 2.0, we extend {\tt ranhel} to be 4, which is at the same level of performing Monte Carlo over the helicity configuration with {\tt ranhel}=3. Instead of  using  $\int_0^{2\pi}{{\rm d}\phi\epsilon_{\phi}^{\mu}(\epsilon_{\phi}^{\nu})^*)}$ to perform the helicity summation where $\epsilon^{\mu}_{\phi}=\sum_{\lambda=\pm,0}{e^{i\lambda\phi}\epsilon^{\mu}}$, we select the helicity eigenstate of external particle when {\tt ranhel}=4 to take a subsequent spin-entangled decay.
\item {\tt pdf} is the PDF set number proposed in
LHAPDF~\cite{Whalley:2005nh}  or in \textbf{pdf}/\textbf{pdflist.txt}. 
Entering $0$ means no PDF is convoluted. If one wants to use LHAPDF, please edit \textbf{input}/\textbf{ho\_configuration.txt} and set the parameter {\tt lhapdf} to be T.
\item {\tt reweight\_pdf} is a flag to use reweighting method to get PDF uncertainty. It only works when using LHAPDF.Correspondingly, one should also specify the first ({\tt pdf\_min}) and the last ({\tt pdf\_max}) of the error PDF sets.
\item {\tt reweight\_Scale} is a flag to use reweighting method to get renormalization and factorization scale dependence, which requires {\tt alphasrun}=T. One can change the lower bound and upper bound for renormalization/factorization scale variations via parameters {\tt rw\_RScale\_down},{\tt rw\_RScale\_up},{\tt rw\_FScale\_down} and {\tt rw\_FScale\_up}.
\item {\tt useMCFMrun} is a flag to perform the strong coupling $\alpha_S$ renormalization group running in the MCFM~\cite{Campbell:2010ff} way.
\item {\tt toodrawer\_output}, {\tt gnuplot\_output}, {\tt root\_output} are flags to ask \HELACOnia\
 to plot histograms and to output into \TopDrawer, \Gnuplot\ and \Root\ files on the fly.
\item {\tt emep\_ISR\_shower} is a parameter to determine whether use \QEDPS\ to take into account initial state radiation effects in electron-positron collisions (1) or not (0).
\item {\tt parton\_shower} is a parameter to determine whether perform parton shower. {\tt parton\_shower}=0 means no shower, i.e. fixed-order calculation. The shower can only be used when the corresponding parton shower program is already installed and the user has already edited properly in \textbf{input}/\textbf{ho\_configuration.txt}. 
\end{enumerate}
All other parameters are listed in {\tt default.inp}. The user can
fix his/her values in {\tt user.inp} following the format in {\tt
default.inp}.

\section{Addon codes\label{sec:addon}}

In this section, we will describe some addon codes implemented in \HELACOnia\ 2.0 for dedicated studies. All of the addon codes have been listed in \textbf{addon}/\textbf{addon\_process.dat}.

\subsection{Single-quarkonium hadroproduction with crystal ball function\label{subsec:sq}}

In fact, the description of the quarkonium-production mechanisms is still a challenge for theorists, especially current the state-of-the-art computation in NRQCD cannot describe the single-quarkonium-hadroproduction data in the whole kinematical region. It would be quite interesting and might be necessary to use emiprical function to describe the single-quarkonium production from $pp$ or $p\bar{p}$ collisions with a data-driven way and use it to test other mechanisms like DPS or $pA$ and $AA$ collisions. Moreover, it also provides an economy way to generate events for single-quarkonium hadroproduction. Therefore, \HELACOnia\ 2.0 has already been implemented some dedicated codes to fit the single-quarkonium-hadroproduction data and to generate events of the single-quarkonium hadroproduction.

Let us start with the description of the calculation. The initial-averaged squared amplitude for single-quarkonium $\mathcal{Q}$ hadroproduction with the assumption of the dominance of gluon-gluon channel can be expressed in a crystal ball function~\cite{Kom:2011bd}
\bq
\overline{|\mathcal{A}_{gg\to\mathcal{Q}+X}|^2}=\left\{
\begin{array}{ll}
K\exp(-\kappa\frac{P_T^2}{M_{\mathcal{Q}}^2})
&\mbox{when $P_T\leq \langle P_T\rangle$} \\
K\exp(-\kappa\frac{\langle P_T \rangle^2}{M_{\mathcal{Q}}^2})\left(1+\frac{\kappa}{n}\frac{P_T^2-\langle P_T \rangle^2}{M_{\mathcal{Q}}^2}\right)^{-n}
&\mbox{when $P_T> \langle P_T\rangle$} \\
\end{array}
\right.
\eq
where $K=\lambda^2\kappa\hat{s}/M_{\mathcal{Q}}^2$ and $\hat{s}$ is the partonic center-of-mass energy. Then the cross section of single quarkonium $\mathcal{Q}$ production in $pp$ collisions is
\bq
\sigma(pp\to \mathcal{Q}+X)=\int{{\rm d}x_1{\rm d}x_2 f_g(x_1)f_g(x_2)\frac{1}{2\hat{s}}\overline{|\mathcal{A}_{gg\to\mathcal{Q}+X}|^2}{\rm dLIPS}},
\eq
where $f_g$ is the gluon PDF and ${\rm dLIPS}$ is the Lorentz-invariant phase space measure for $pp\to \mathcal{Q}+X$. The coefficients $\lambda$,$\kappa$,$n$ and $\langle P_T\rangle$ can be determined by fitting it to the experimental data.

\subsubsection{Fit codes}
The codes for fitting $\psi(1S,2S)$ and $\Upsilon(1S,2S,3S)$ are in subdirectories \textbf{fit\_pp\_psiX\_CrystalBall} and \textbf{fit\_pp\_upsilonX\_CrystalBall} respectively. One can use the command lines 
\cCode{}
HO> generate addon 3
\end{lstlisting}
and 
\cCode{}
HO> generate addon 4
\end{lstlisting}
to drive the corresponding codes to perform fitting with \Minuit~\cite{James:1975dr} package. The input files dedicated to these {\it ad hoc} codes are in \textbf{fit\_pp\_psiX\_CrystalBall}/\textbf{input} 
and \\
\textbf{fit\_pp\_upsilonX\_CrystalBall}/\textbf{input}. One can specify the meson in \textbf{state.inp} and the fitting parameters in \textbf{fit\_param\_card.inp}. The selected experimental data can be assigned in \textbf{data\_list\_i.inp}, where \textbf{i} is the number in \textbf{state.inp}. Some fitted results are contained in \textbf{fit\_pp\_psiX\_CrystalBall}/\textbf{fitresults} and \textbf{fit\_pp\_upsilonX\_CrystalBall}/\textbf{fitresults}. We have checked the fitted results of Ref.~\cite{Kom:2011bd} for prompt $J/\psi$ production with the same setup.

For instance, through a combined fit of $d^2\sigma/dP_Tdy$ to the ATLAS~\cite{Aad:2014fpa}, CMS~\cite{Chatrchyan:2011kc}, LHCb~\cite{Aaij:2012ag} and CDF~\cite{Aaltonen:2009dm} data, we obtained $\kappa=0.543$ and $\lambda=0.118$ for prompt $\psi(2S)$ when $\langle P_T\rangle = 4.5~\rm{GeV}$ and $n=2$, where $\chi^2=242$ for total $90$ experimental data. The comparisons are shown in Fig.\ref{fig:fit}. The result is collected in \textbf{fitresults}/\textbf{psi2s}/\textbf{fit1}.

\begin{figure}
\begin{center}
\vspace{-2cm}\hspace{-1cm}\includegraphics[width=0.5\textwidth]{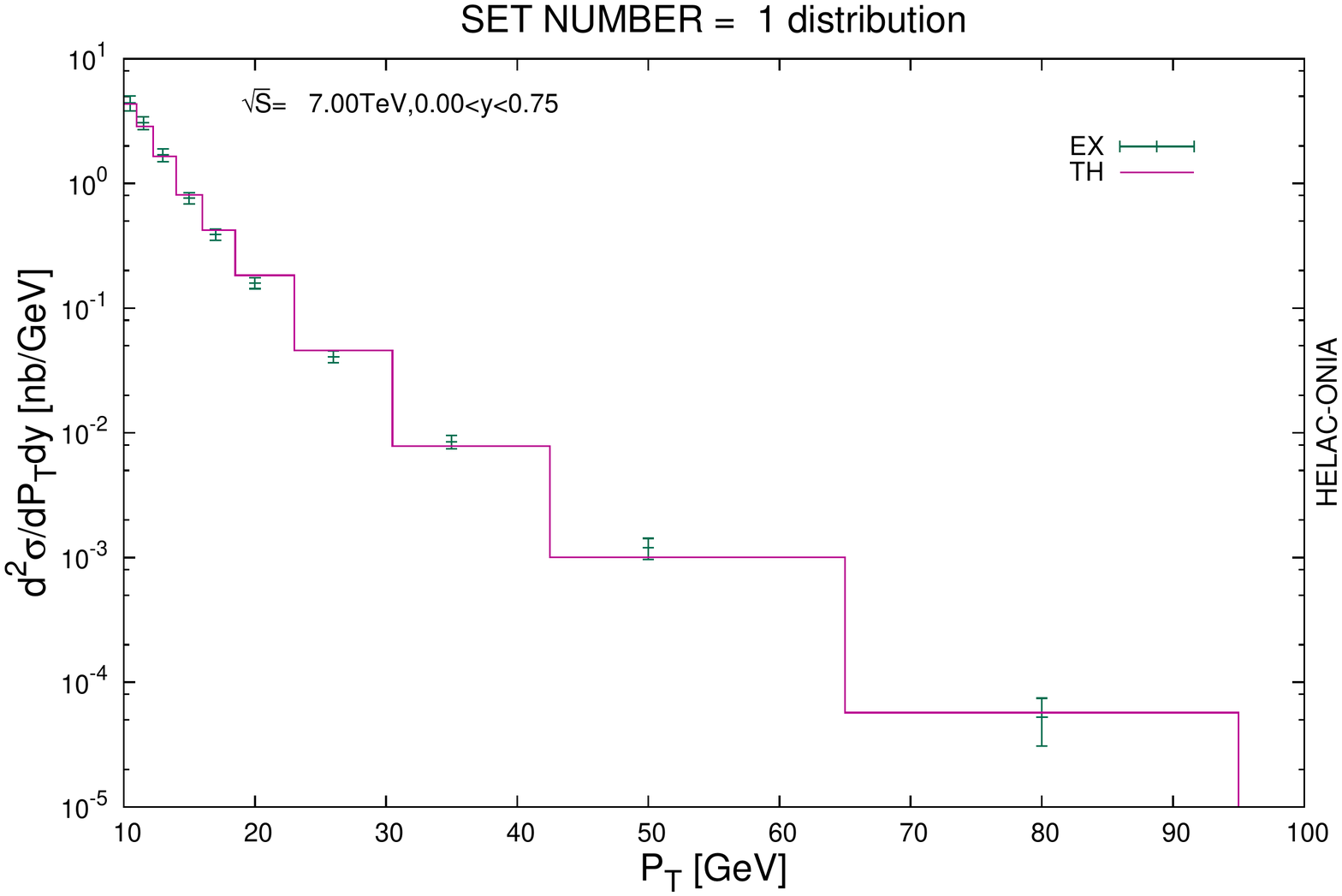}
\hspace{-0.75cm}\includegraphics[width=0.5\textwidth]{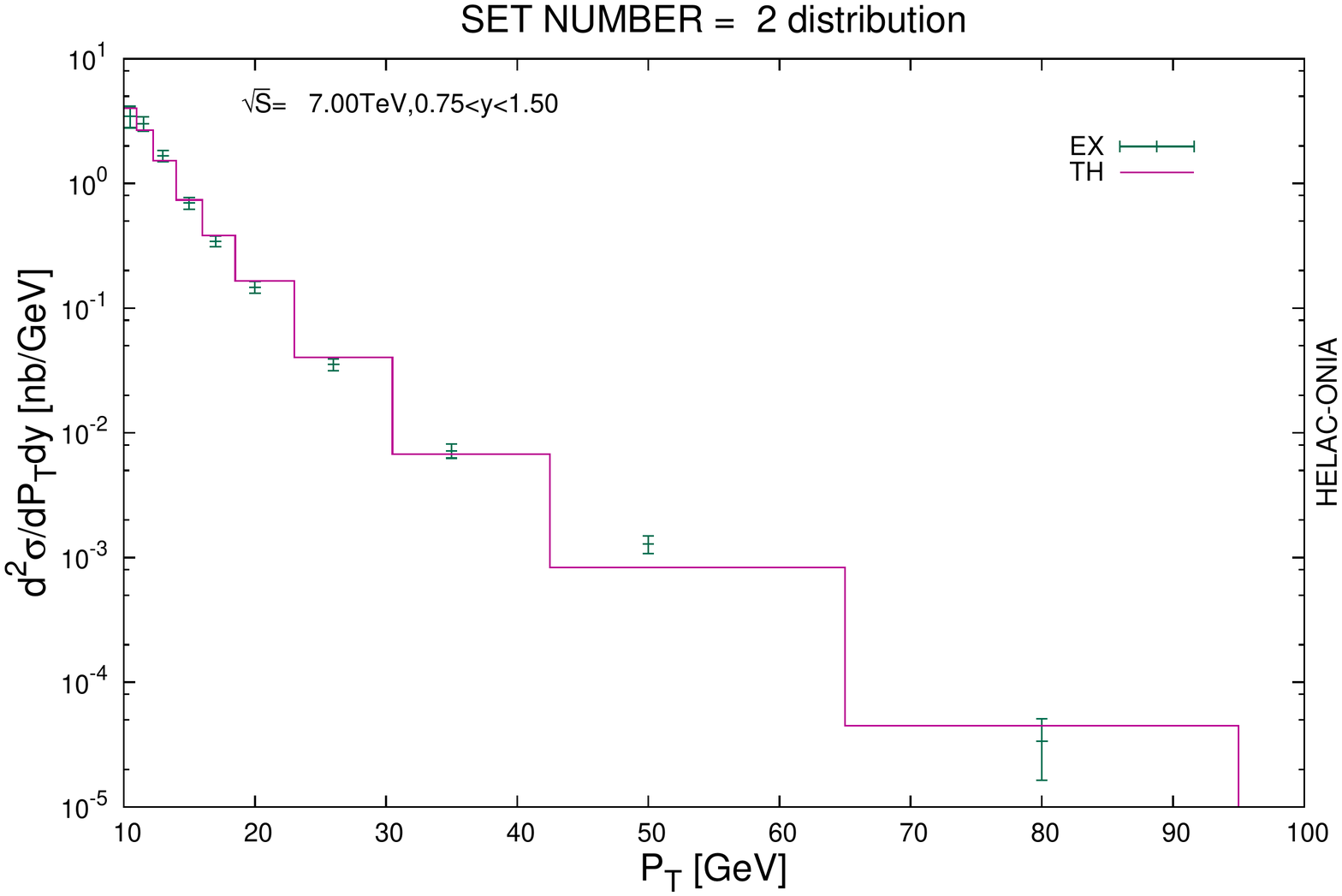}\\
\vspace{-1cm}\hspace{-1cm}\includegraphics[width=0.5\textwidth]{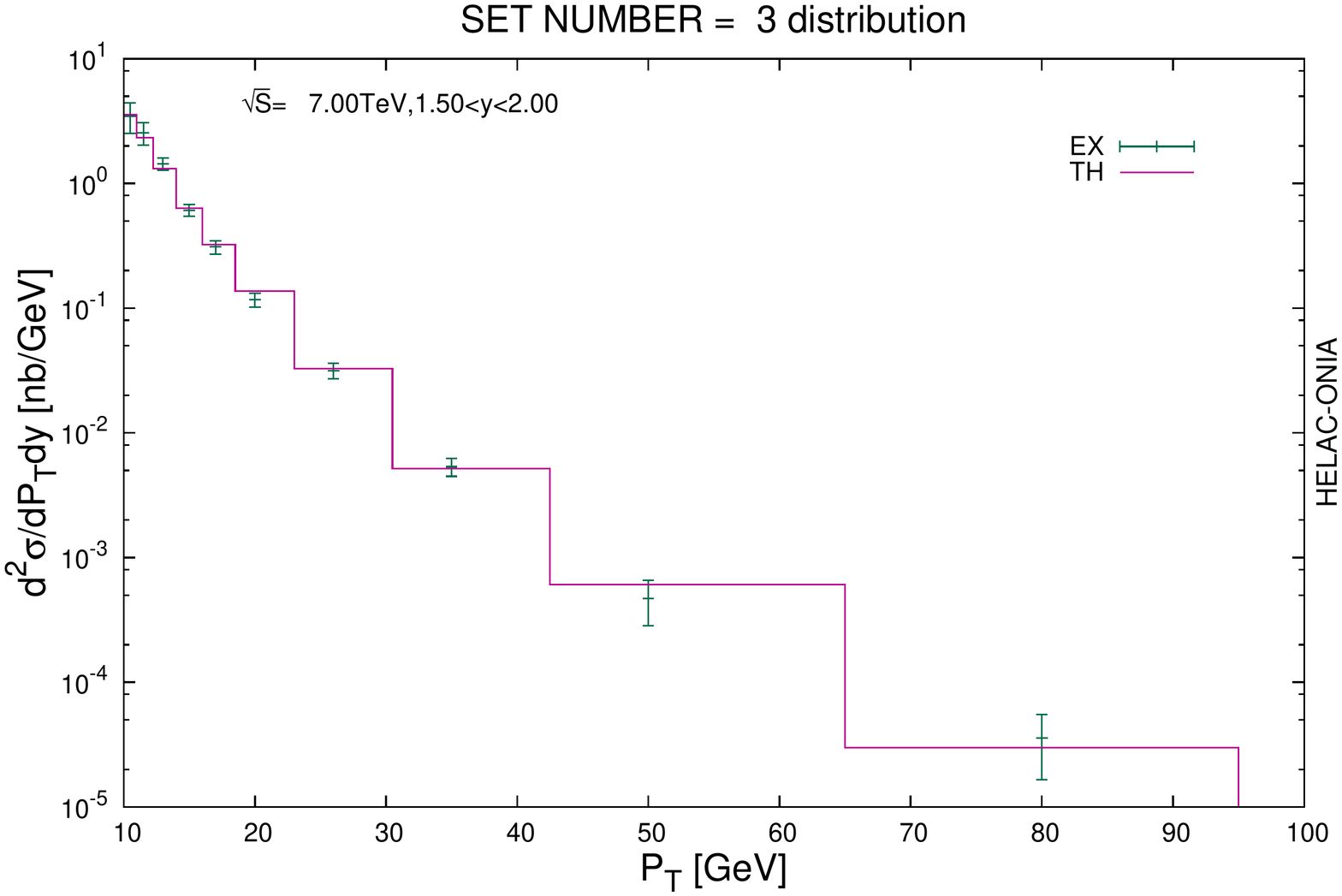}
\hspace{-0.75cm}\includegraphics[width=0.5\textwidth]{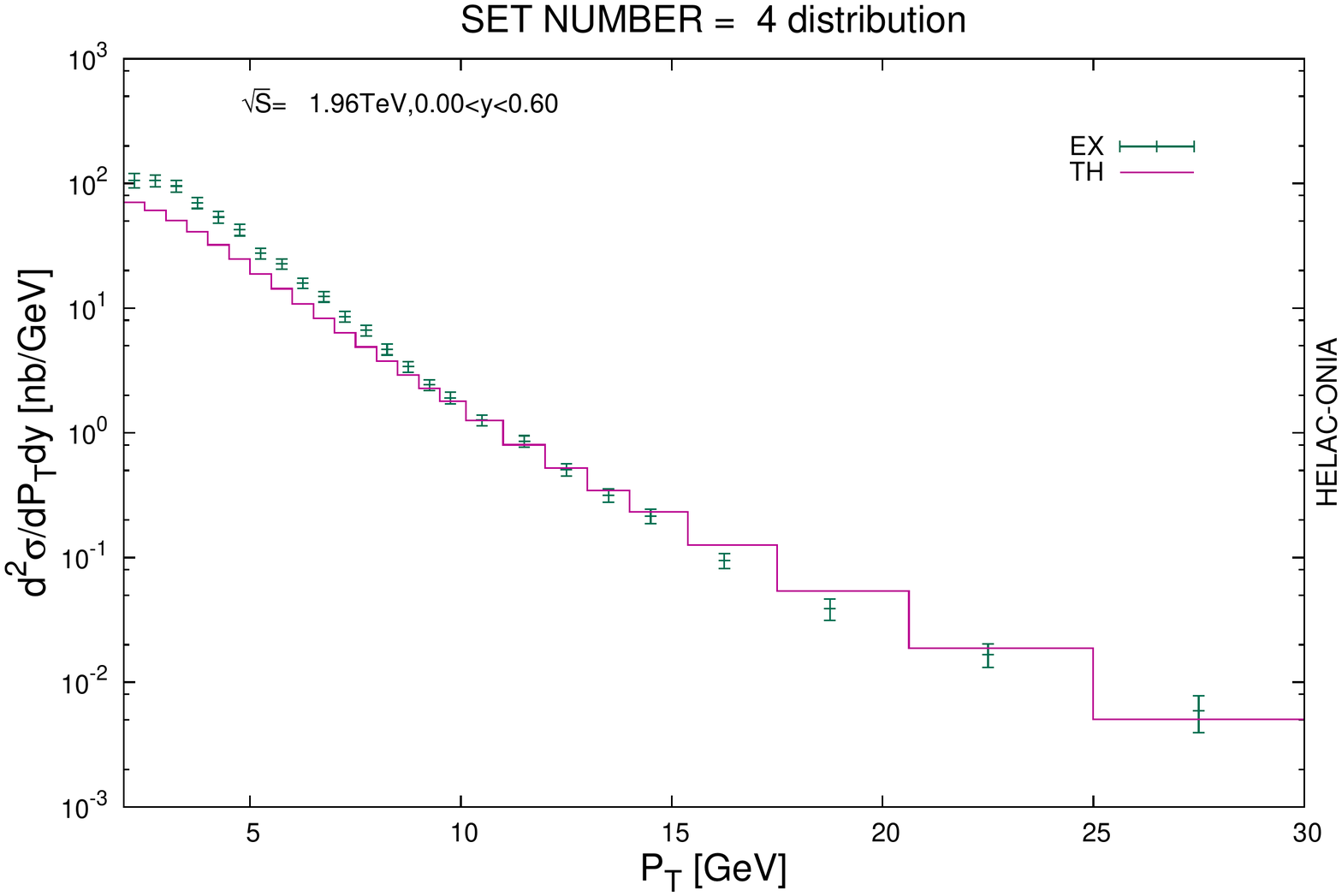}\\
\vspace{-1cm}\hspace{-1cm}\includegraphics[width=0.5\textwidth]{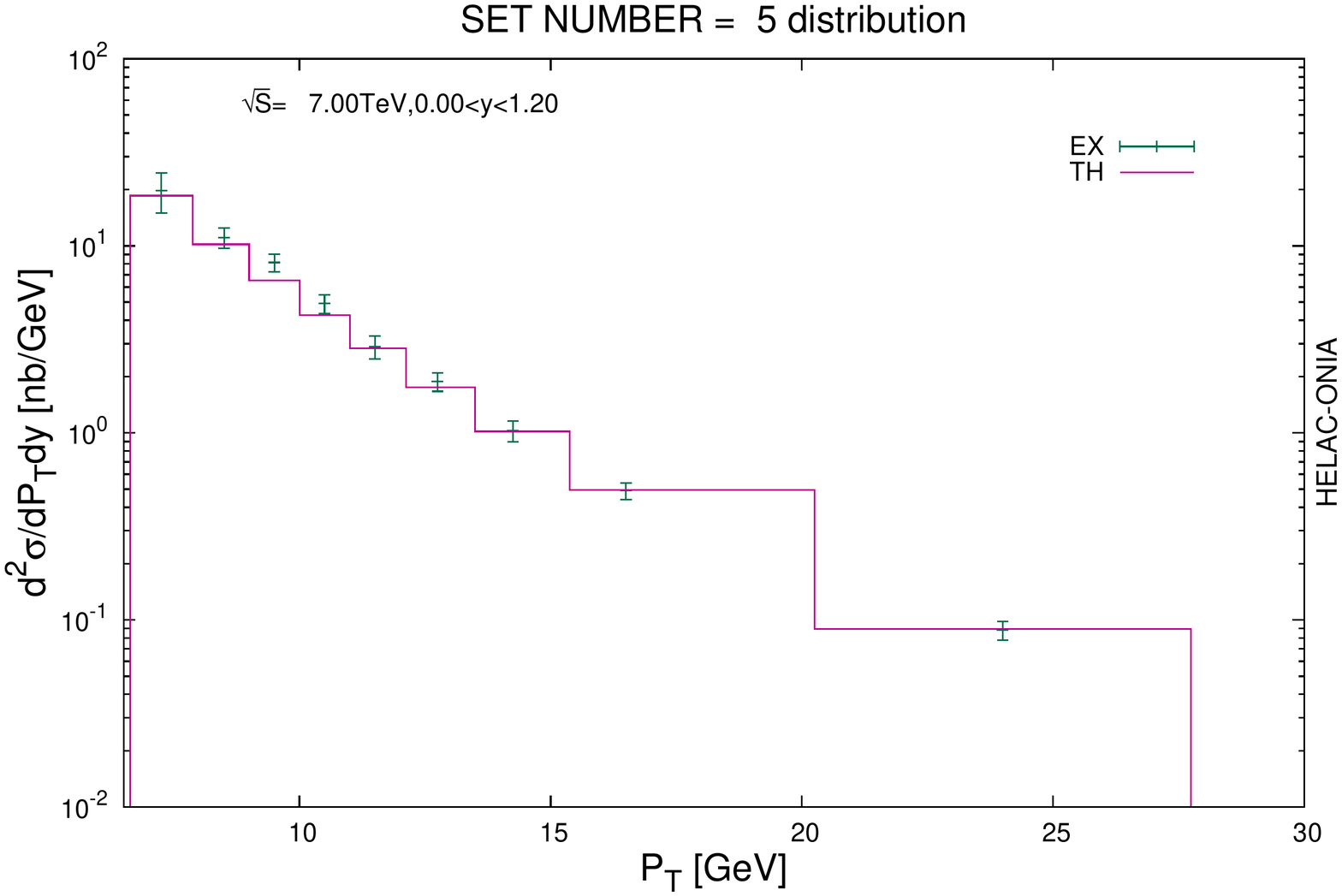}
\hspace{-0.75cm}\includegraphics[width=0.5\textwidth]{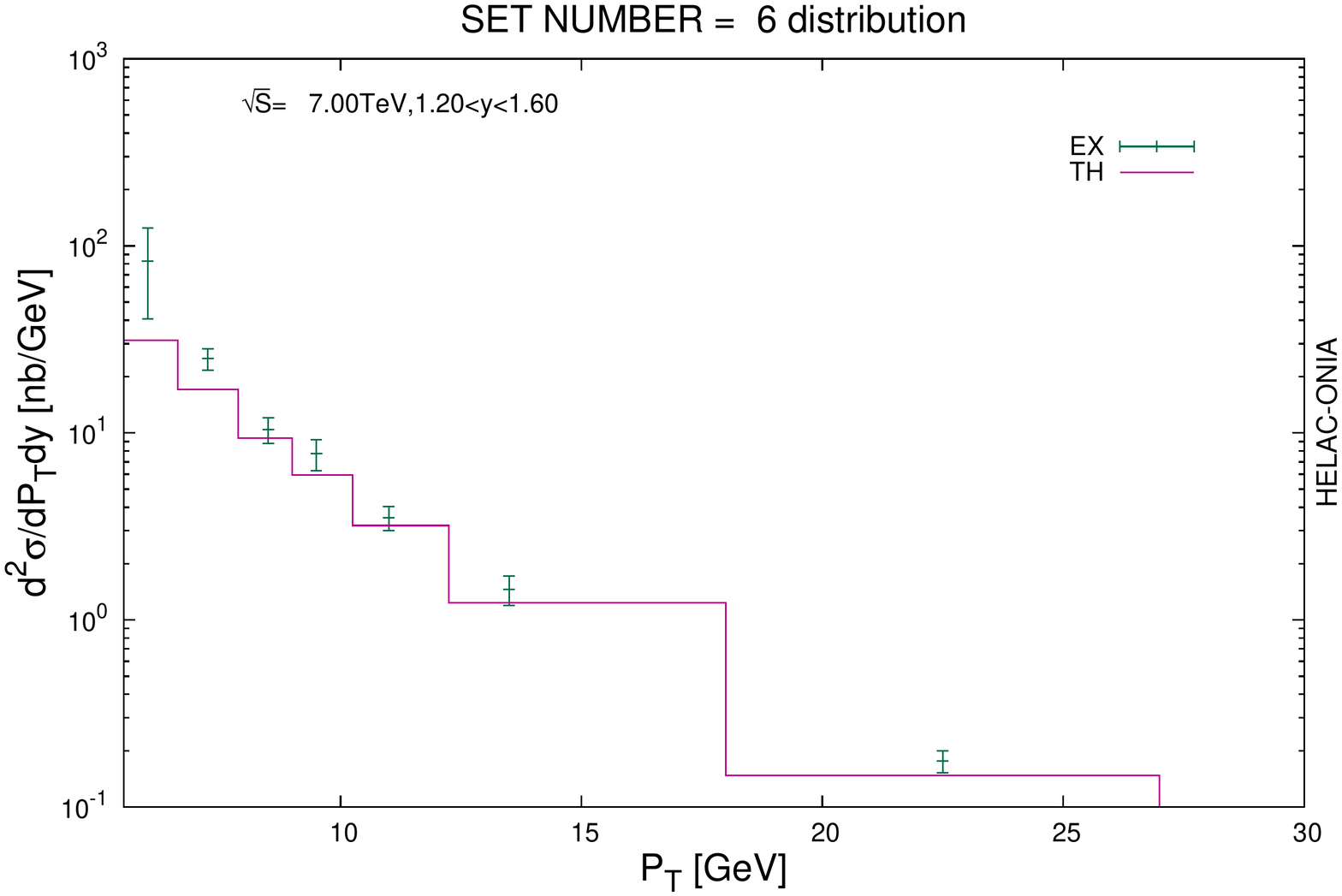}\\
\vspace{-1cm}\hspace{-1cm}\includegraphics[width=0.5\textwidth]{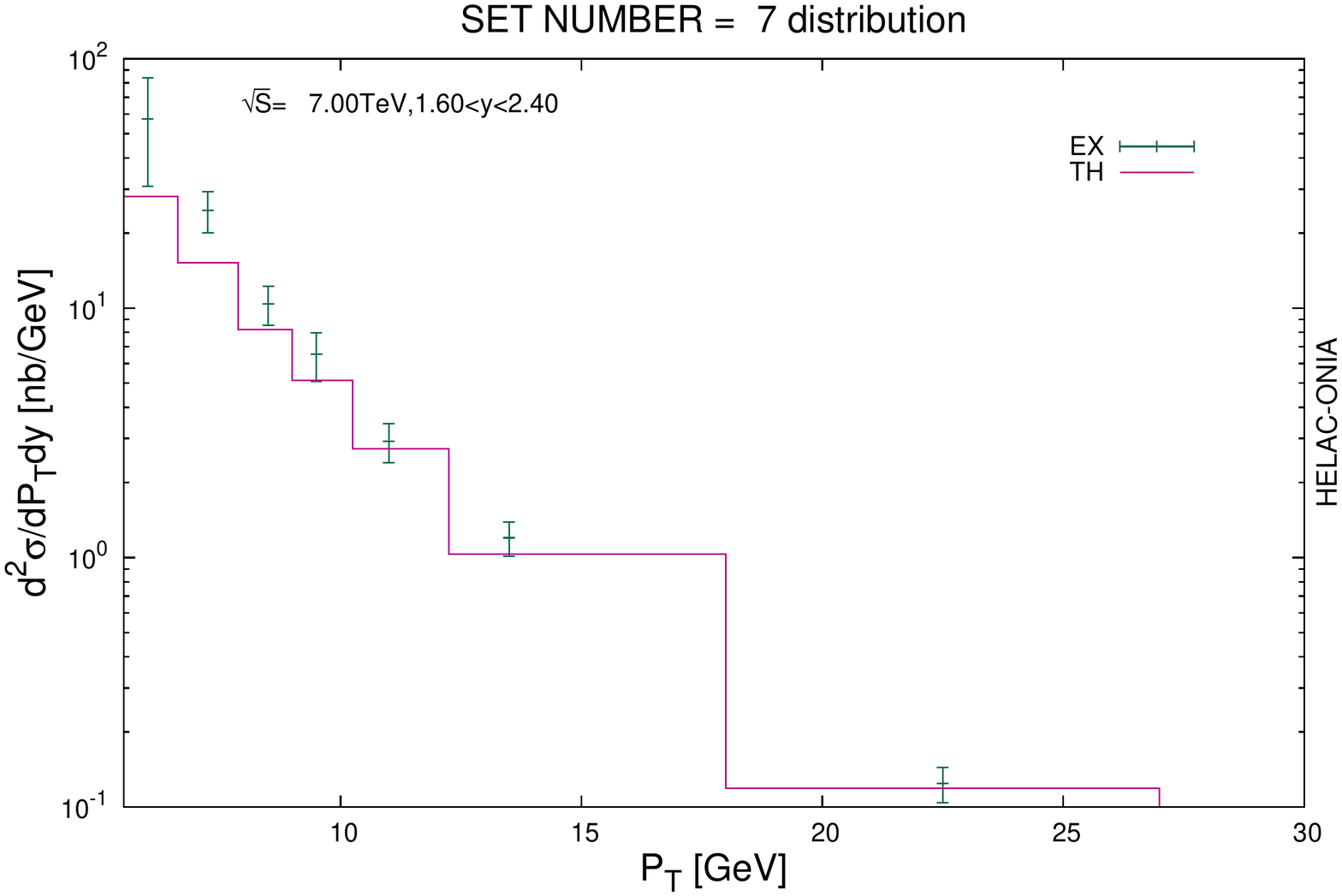}
\hspace{-0.75cm}\includegraphics[width=0.5\textwidth]{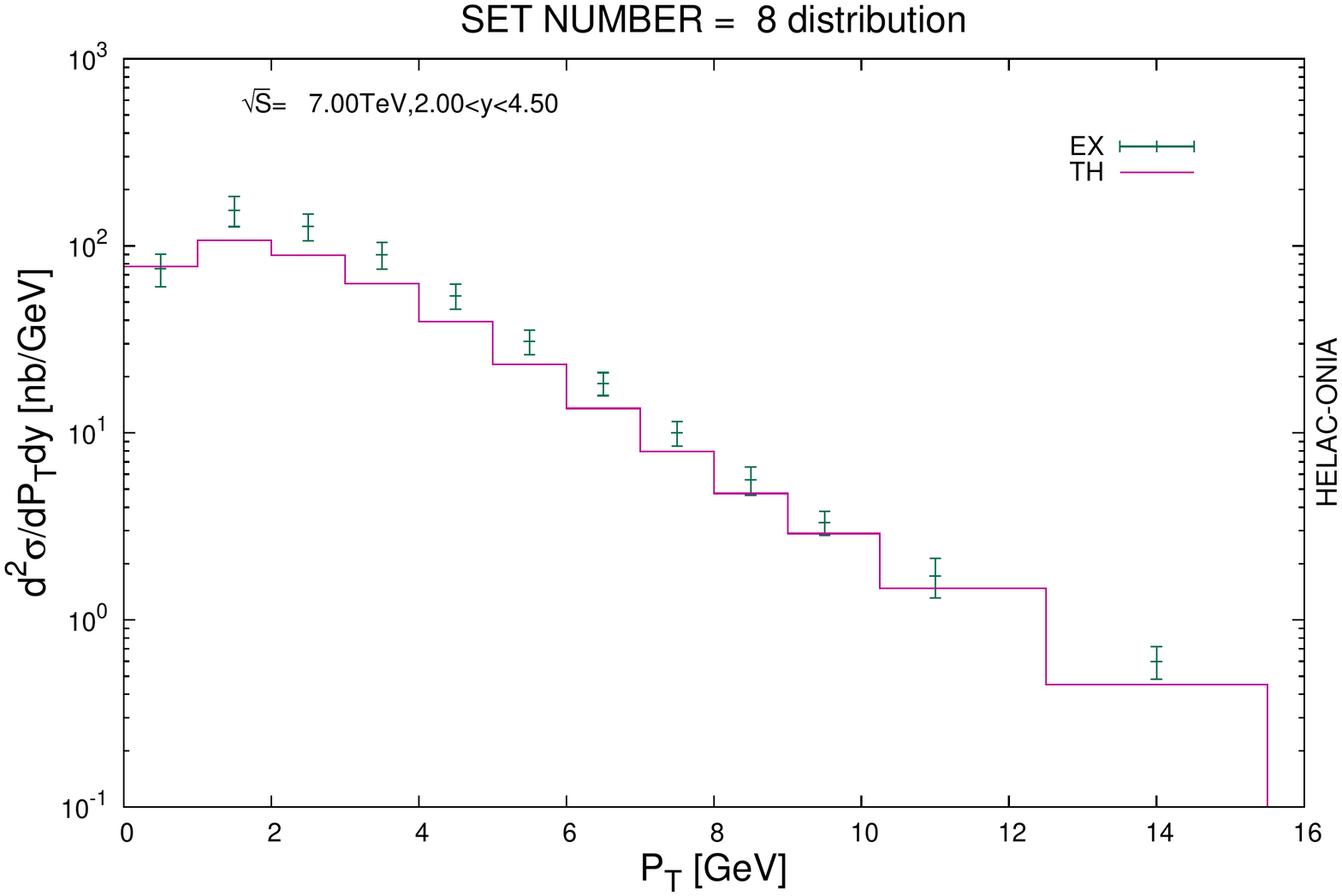}
\vspace{-1cm}
\caption{\label{fig:fit}Combined fit of $d^2\sigma/dP_Tdy$ to ATLAS~\cite{Aad:2014fpa} (1st-3rd plots), CDF~\cite{Aaltonen:2009dm} (4th plot),  CMS~\cite{Chatrchyan:2011kc} (5th-7th plots), LHCb~\cite{Aaij:2012ag} (8th plot) for prompt $\psi(2S)$ production. The plots are generated automatically by \HELACOnia\ 2.0.}
\end{center}
\end{figure}

\subsubsection{A simple event generator for single-quarkonium hadroproduction}

With the fitted parameters, we wrote a simple event generator for $pp(\bar{p})\to\mathcal{Q}+X$, where $\mathcal{Q}=J/\psi,\psi(2S),\Upsilon(1S),\Upsilon(2S),\Upsilon(3S),\chi_{c0},\chi_{c1},\chi_{c2}$ and $\chi_{bJ}(nP)$ with $J=0,1,2$,$n=1,2,3$. The code is put in the subdirectory \textbf{addon}/\textbf{pp\_psiX\_CrystalBall}. One can drive such program with the following command line
\cCode{}
HO> generate addon 2
\end{lstlisting}
Some special input parameters can be specified in \textbf{pp\_psiX\_CrystalBall}/\textbf{input}. One can set the type of $\mathcal{Q}$ in \textbf{state.inp} and its polarization in \textbf{polarization.inp}. The file \textbf{crystalball.inp} is used to input the parameters $\lambda$,$\kappa$,$n$ and $\langle P_T\rangle$. We have performed some applications in Ref.~\cite{Massacrier:2015qba};

\subsection{Double parton scattering for double-quarkonium production}

We also implemented the code for calculating DPS for double-quarkonium production in a $pp$ or $p\bar{p}$ collider. One of its application can be seen in Refs.~\cite{Lansberg:2014swa,Lansberg:2015lva}. We used a simple but widely-used ``pocket formula" to describe DPS for double-quarkonium production $pp\to\mathcal{Q}_1\mathcal{Q}_2+X$
\bq
\sigma^{\rm DPS}_{\mathcal{Q}_1\mathcal{Q}_2}=\frac{1}{1+\delta_{\mathcal{Q}_1\mathcal{Q}_2}}\frac{\sigma_{\mathcal{Q}_1}\sigma_{\mathcal{Q}_2}}{\sigma_{\rm eff}},\label{eq:DPSQQ}
\eq
where $\sigma_{\mathcal{Q}_i}$ is the cross section for single quarkonium $\mathcal{Q}_i$ production and $\sigma_{\rm eff}$ is a parameter to characterise an effective spatial area of the parton-parton interactions. $\sigma_{\rm eff}$ can be related to the parton spatial density $F(\bf{b})$ inside the proton as 
\bq
\sigma_{\rm eff}=\left[\int{{\rm d}^2\bf{b}\left(F(\bf{b})\right)^2}\right]^{-1}.
\eq
Within the factorization, $\sigma_{\rm eff}$ should be independent of final states but it might change with different species of initial partons and its Bjorken fraction $x$. A first order assumptions of $\sigma_{\rm eff}$ is independent of process and energy, which however should be checked case by case.

For single-quarkonium production, we used the crystal ball function described in appendix \ref{subsec:sq} to estimate the squared amplitude. This special code can be found in \textbf{addon}/\textbf{pp\_psipsi\_DPS}. The command line to generate this process is
\cCode{}
HO> generate addon 1
\end{lstlisting}
Similar to other addon codes, the input parameters dedicated to this code is in \textbf{pp\_psipsi\_DPS}/\textbf{input}. The parameter $\sigma_{\rm eff}$ can be specified in \textbf{sigma\_eff.inp}. The user can change the type of the quarkonium pair $\mathcal{Q}_1$ and $\mathcal{Q}_2$ in the file \textbf{states.inp}. The polarizations of $\mathcal{Q}_i$ can be fixed in \textbf{polarization\_{\ttfamily{<name>}}.inp}, where {\ttfamily{<name>}} is the name of $\mathcal{Q}_i$,i.e. $J/\psi=\textbf{jpsi},\psi(2S)=\textbf{psi2S},\Upsilon(1S)=\textbf{Y1S},\Upsilon(2S)=\textbf{Y2S},\Upsilon(3S)=\textbf{Y3S}$. The parameters $\lambda$,$\kappa$,$n$ and $\langle P_T\rangle$ in the crystal ball function should be told in the files \textbf{crystalball\_{\ttfamily{<name>}}.inp}.

\subsection{Heavy-flavor quark pair hadroproduction with crystal ball function}

Using the crystal ball function for heavy-flavor quark pair or open heavy-flavor meson pair production, one can also perform a fit to experimental data at the hadronic colliders in the  $P_T$ spectrum of the quark/meson production. Hence, it also provides an opportunity to use a data-driven way to analyze the corresponding heavy-flavor quark/meson pair production, which usually suffers large theoretical uncertaities in a perturbative computation. Such a method indeed has been applied in the open charm production at a proposed fixed-target experiment at the LHC (AFTER@LHC) in Ref.~\cite{Massacrier:2015qba}. The fit can be performed using the following commands
\cCode{}
HO> generate addon 5
\end{lstlisting}
Some input parameters for fitting are needed to be specified in \textbf{fit\_pp\_QQ\_CrystalBall}/\textbf{input}. Moreover, with the fitted parameters, one can use the 
\cCode{}
HO> generate addon 6
\end{lstlisting}
to generate the unweighted events for the heavy quark pair production in proton-proton or proton-antiproton collisions.

\subsection{Double parton scattering for associated production of dipoton and dijet}

Similar to Eq.(\ref{eq:DPSQQ}), we have a ``pocket" formula for diphoton and dijet production via DPS mechanism in $pp$ collisions
\bqa
\sigma^{\rm DPS}_{\gamma\gamma+jj}=\frac{\sigma_{\gamma\gamma}\sigma_{jj}}{\sigma_{\rm eff}}+\frac{\sigma_{\gamma+j}\sigma_{\gamma+j}}{2\sigma_{\rm eff}}.
\eqa
The LO matrix elements of $ab\to\gamma+\gamma,\gamma+j,j+j$ have been implemented in \HELACOnia\ with the correct color flow. To the diphoton production, we also implemented the gluon-gluon initial state process, which is a loop-induced process for diphoton production. However, due to the high luminosity of the gluon-gluon initial state at a high energy collider, such a contribution might be substantial. Unweighted events for the DPS contributions to diphoton and dijet production can be generated by the command
\cCode{}
HO> generate addon 7
\end{lstlisting}
One can change the input $\sigma_{\rm eff}$ in the input file \textbf{pp\_aajj\_DPS}/\textbf{input}/\textbf{sigma\_eff.inp}. There is also a file \textbf{pp\_aajj\_DPS}/\textbf{input}/\textbf{subprocess.inp} for user to select/drop some partonic subprocesses and to choose to generate the unweighted events of $pp\to\gamma+\gamma,pp\to \gamma+j,pp\to j+j$ instead of the DPS process. Such a functionality is very useful to cross check and to specify a global K-factor from the missing higher-order quantum corrections. Some studies on the inclusive DPS production rates of this process at the Tevatron~\cite{Drees:1996rw} and the LHC~\cite{Tao:2015nra} have been explored in the literature.

\vfill

\noindent{\Large \bf Acknowledgements} \\[24pt]
I thank Jean-Philippe Lansberg for motivating me to improve the tool by its applications on several relevant physics projects, and for proofreading the manuscript. I am grateful to Chole Gray, Darren Price, Barbara Trzeciak for the feedback on using the program. Finally, I would also like to thank the authors of \MG5aMC\ to collaborate on the amazing \MG5aMC\ project, from which I indeed learned a lot on how to improve the user friendliness of \HELACOnia. This work was supported by ERC grant 291377 ``LHCtheory: Theoretical predictions and analyses of LHC physics: advancing the precision frontier".

\vspace*{2cm}


\providecommand{\href}[2]{#2}\begingroup\raggedright\endgroup

\end{document}